\documentclass[journal]{IEEEtran}

\usepackage{amsmath, amsfonts}
\usepackage{algorithmic}
\usepackage{graphicx}
\usepackage{textcomp}
\usepackage{hyperref}
\usepackage{tikz}
\usepackage{caption}
\usepackage{subcaption} 
\usepackage{float}
\usepackage{flushend}
\usepackage{multirow}
\usepackage{enumitem}
\usepackage{tablefootnote}
\usepackage{booktabs}
\usepackage{threeparttable}
\usepackage{adjustbox}
\usepackage{tabularx}
\usepackage{array}
\usepackage{ragged2e}
\usepackage[linesnumbered,ruled]{algorithm2e}
\usepackage{boldline} 
\usepackage{makecell}
\usepackage{accents}
\usepackage{cite}
\usepackage{url}
\usepackage{textcomp}
\usepackage{stfloats}
\usepackage{orcidlink}

\newcommand{\boldhline}{\Xhline{2.5\arrayrulewidth}}

\usepackage{xcolor}
\def\BibTeX{{\rm B\kern-.05em{\sc i\kern-.025em b}\kern-.08em
    T\kern-.1667em\lower.7ex\hbox{E}\kern-.125emX}}

\def\checkmark{\tikz\fill[scale=0.4](0,.35) -- (.25,0) -- (1,.7) -- (.25,.15) -- cycle;}

\newcommand{\sour}[1]{\textcolor{black}{ #1}} 

\begin{document}

\title{Base Station Certificate and Multi-Factor Authentication for Cellular Radio Control Communication Security}

\author{\IEEEauthorblockN{Sourav Purification\IEEEauthorrefmark{1},
Simeon Wuthier\IEEEauthorrefmark{1},
Jinoh Kim\IEEEauthorrefmark{2}, 
Ikkyun Kim\IEEEauthorrefmark{3}, and
Sang-Yoon Chang\IEEEauthorrefmark{1}} \\
\IEEEauthorblockA{\IEEEauthorrefmark{1}
University of Colorado Colorado Springs, \textit{\{spurific, swuthier, schang2\}@uccs.edu}}
\\
\IEEEauthorblockA{\IEEEauthorrefmark{2}
Texas A\&M University-Commerce, \textit{jinoh.kim@tamuc.edu}}
\\
\IEEEauthorblockA{\IEEEauthorrefmark{3}Electronics and Telecommunications Research Institute, \textit{ikkim21@etri.re.kr}}
}





\maketitle

\begin{abstract}


Current cellular networking remains vulnerable to malicious fake base stations due to the lack of base station authentication mechanism or 
even a key to enable authentication. 
We design and build a base station certificate (certifying the base station's public key and location) and a multi-factor authentication (making use of the certificate and the information transmitted in the online radio control communications) to secure the authenticity and message integrity of the base station control communications. 
We advance beyond the state-of-the-art 
research by introducing greater authentication factors (and analyzing their individual security properties and benefits), and by using blockchain 
to deliver the base station digital certificate offline (enabling greater key length/security strength and computational/networking efficiency). 
We 
design the certificate construction, delivery, and the multi-factor authentication use on the user equipment. The user verification involves multiple factors verified through the ledger database, the location sensing (GPS in our implementation), and the cryptographic signature verification of the cellular control communication (SIB1 broadcasting). 
%
We analyze 
our scheme's 
security, performance, and the fit to the existing standardized networking protocols. 
Our work involves the implementation 
building on X.509 certificate (adapted), smart contract-based blockchain, 5G-standardized RRC control communications, and software-defined radios. 
Our 
analyses 
show that our scheme effectively defends against more security threats and can enable stronger security, i.e., ECDSA with greater key lengths. 
Furthermore, our scheme 
enables the computing and energy to be more than three times 
efficient than the previous research on the mobile user equipment. 

\end{abstract}





\section{Introduction}

Mobile and wireless devices use cellular networking, e.g., 4G and 5G, for connectivity and Internet access. Cellular networking has exceeded Wi-Fi in accessing the Internet and web since November 2016. 
As of April 2025, the cellular traffic is 1.77 times greater in use/traffic worldwide and up to 6 times greater in the developing regions~\cite{mbdtus}. 
To access the remote servers, the mobile users rely on the cellular base station which is the first hop in the communication path and acts as a bridge gateway between the wireless link (from/to the mobile user) and the wired connections (from/to the cloud server via routers and switches). 

The malicious or rogue base station has been a well-known security issue in the wireless security community, as described in Section~\ref{sec:related_work}. The malicious base station threats include breaching the user privacy against its location and credentials, redirection of the networking (e.g., fake destination and web server), protocol downgrading and manipulation (e.g., de-registration and authentication-and-key-agreement or AKA bypass), 
dispatching false alert messages, 
and complete control of the availability and wireless link for DoS, among others. 
%
The vulnerability stems from the lack of key establishment and authentication of the base stations\footnote{The 3GPP 5G provider infrastructure does support keys and authentication but the subject entities are at the backend beyond the base station and the security setup occurs after the base station's control communications for wireless MAC and RRC. Specifically, 5G authentication and key agreement (AKA) is between the user equipment and the core network, while 5G authentication and key management for applications (AKMA) is with the destination application servers. 
For example, 5G AKA
establishes the public key for the backend core network, e.g., via hardcoding on the sim card, and there is security set up between the core network and the user equipment~\cite{3gppsap}. However, the intermediary base station does not have a key of its own to build autonomous security control on its communications with the user equipment.}, which prevents a mobile user from distinguishing between authorized vs. rogue/malicious base station. 
There have been recent proposals in research~\cite{singla2021look,hussain2019insecure} and standardization/development~\cite{3gppfbs} to introduce and use the base station's public key. 
In 3GPP standardization, such proposals are at the early stages of establishing the requirements for the base station keys~\cite{3gppfbs}. 
The lack of key disables the authentication and the integrity protection of the control communication exchanged between the user equipment and base station; such control communication resolves the medium access control (MAC) and radio resource control (RRC) to set up the wireless channel before involving the backend core network. 

\begin{figure}[t!]
    \centering
\includegraphics[width=0.9\columnwidth,height=2cm]{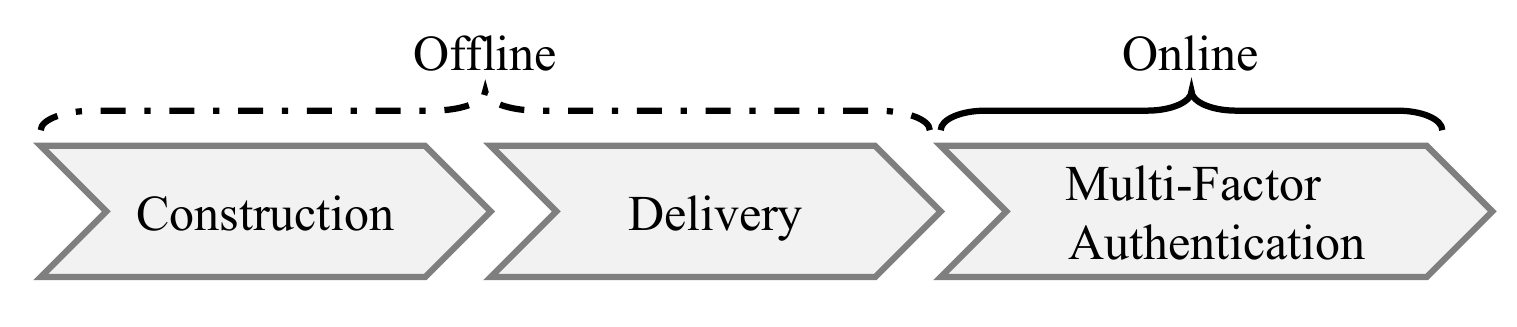}
    \caption{Our contributions spanning across the certificate processes in online vs. offline.} 
    \label{fig:arrow_diagram}
\end{figure}

We introduce and design base station digital public-key certificate, which establishes, binds, and distributes the base station's public key to user equipment accessing the cellular connectivity. 
We design and build our scheme across the base station certificate construction, delivery, and use for multi-factor authentication. These discrete processes are shown in Figure~\ref{fig:arrow_diagram} and include the offline phases (any time before the user equipment wishes to access connectivity and initiates the connectivity request) vs. the online phases. 
Our scheme design across these processes 
are informed by our vulnerability and threat analyses of fake base station. 

Our base station certificate \emph{construction} builds on the popular and standardized X.509 certificate including its information fields. We analyze how these information fields can defend against the various base station threats and add the geographical location information in the certificate, which has not been used in the existing certificates for other applications. The terrestrial base stations are static and has fixed location, and we include their location information in our construction and bind it and the ID to the public key for location authenticity. 
The location information addition on the base station certificate specifically enables additional geographical-sensing-based verification and the defense against wormhole threats. 

Our base station certificate \emph{delivery} shares the certificate with the user equipment so that the user equipment can use it to verify the authenticity when accessing the cellular connectivity provided by the base station. We use blockchain to provide the transparent list of the registered base stations and enable the \emph{offline} delivery of the certificate any time before the \emph{online} when the user equipment accesses the cellular connectivity; we describe offline vs. online in the 5G context and explain how offline is cheaper and provides greater flexibility in Section~\ref{subsec:offline_vs_online}. 
The offline delivery provides energy/computing efficiency to the user equipment. The offline delivery also enables the application of the cryptographic cipher with greater key length/security than the state-of-the-art research (which delivers the key in online) due to the 3GPP-standardized protocol limiting the control message size (i.e., the base station's SIB1 message is limited to 372 Bytes as described in Section~\ref{subsec:background_5G}).  

We then use the offline-delivered base station certificate to enable the \emph{multi-factor authentication} of the base station in online control communication. 
The user equipment verification of the subject base station is sequential and uses the multiple factors of ID, $L$, $t$, and the digitally signed SIB1 broadcasting message. These factors are verified and cross-checked with multiple sources, from the stored ledger to the location sensing to the public-key verification.  
First, the user equipment uses the ID that enables it to look up its locally stored blockchain ledger to verify that the base station is a legitimate registered base station. Second, it then accesses the base station's registered location information, $L$, stored in the local ledger, and cross-checks with its sensed location to verify the physical location integrity. Third, it verifies the freshness of the online broadcasting control communication of SIB1 
using the timestamp $t$. 
Fourth, it verifies the message authenticity of the digitally signed SIB1 by using the base station's public key, $K$ to verify the signature. 
The subject base station is authenticated if all the factors are authenticated and verified. 

Our scheme thus not only verifies the base station but also verifies the message authenticity of the SIB1 control communication message. 
The message authenticity protection for the control communications is critical as the \emph{control communications} are the communication exchanges to set up the communication channels for the following data communications or connectivity service, e.g., radio channel frequency and medium access control (MAC) with the base station. The vulnerability in such control communication can enable handover exploitation, dispatching false alert messages, and bidding down threats, as described in Section~\ref{sec:related_work}.  
In 4G and 5G, such control communication is called radio resource control (RRC), and we digitally sign the periodic system information block 1 (SIB1) broadcasting message announcing the base station's presence and starting RRC. 

We take a systems approach to make contributions across multiple process steps for authenticating the base station's ID and the control communication. 
We conduct comprehensive analyses of the vulnerability and the threat based on a fake/malicious base station in 5G networking to inform our scheme design and identify how our scheme defends against threats based on a malicious or rogue base station. 
Our research also involves prototype implementation for validation and experimentation, and builds on the existing standardized technologies. 
For the offline processes, 
we build on the X.509 certificate and use the blockchain to deliver the base station's public key and location. Blockchain automatically delivers and synchronizes the latest public key information when the core network introduces/adds the subject base station to the connectivity-provision infrastructure, and we use an Ethereum smart contract in our implementation. 
For the online processes, 
we build on the 3GPP 5G NR protocol and more precisely the SIB1 message broadcasting transmissions in the 5G RRC. 
The user equipment, having received the base station's certificate and public key previously in offline, can verify the base station's authenticity via multi-factor authentication, including verifying the digitally signed SIB1 message. We implement, validate, and test our scheme using software-defined radios (SDR) and the open-source 4G/5G srsRAN software. 

To highlight our contributions, we analyze the vulnerability, threats, and the existing defense research in the base station in 5G networking ($\S$\ref{sec:threat_analyses}). Informed by such vulnerability/threat analyses, we design our base station certificate scheme spanning across construction, delivery, and use for multi-factor authentication ($\S$\ref{sec:our_scheme}); both the certificate construction of the information fields and the offline blockchain-based delivery enables the multi-factor authentication and advances its security and performance beyond the existing practice and research. 
We implement our scheme using SDR and various computing platforms for hardware, and srsRAN and Ethereum smart contract for software ($\S$\ref{sec:experimental_setup}). 
We use implementation to experimentally validate our scheme for proof-of-concept and for empirical security and performance analyses ($\S$\ref{sec:experimental_analyses}). 
We compare our scheme with the state-of-the-art research in design and approaches ($\S$\ref{sec:related_works}) 
and in empirical validation ($\S$\ref{subsec:security_analysis},$\S$\ref{subsec:results_online}) to show that our scheme defends greater threats including the wormhole threat, enables digital signature with longer key length, and the time and energy are more than three times efficient at the user equipment.

\begin{figure*}
    \centering
    \begin{subfigure}{1\columnwidth}
      \centering
      \includegraphics[width=0.98\textwidth]{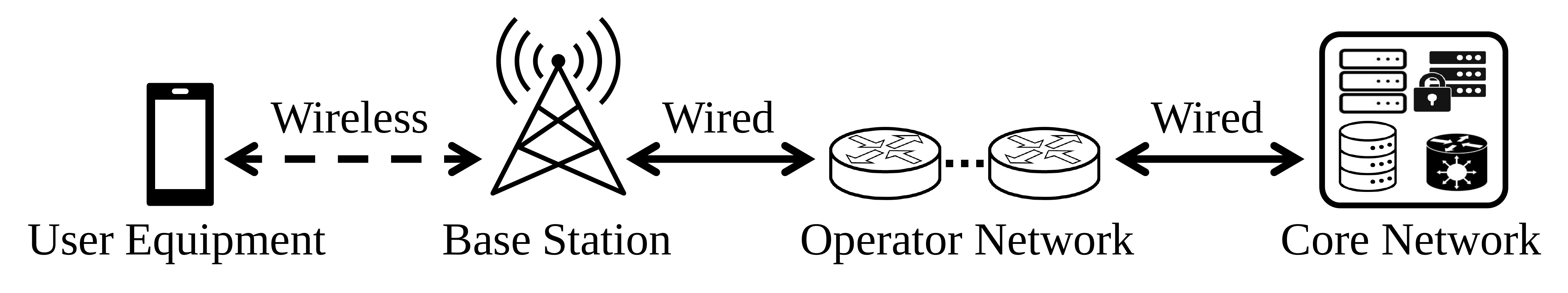}
      \caption{Architecture including user equipment, base station, and core network.}
      \label{fig:physical_arch}
    \end{subfigure}
\hfill
    \begin{subfigure}[b]{1\columnwidth}
      \centering
      \includegraphics[width=0.85\textwidth]{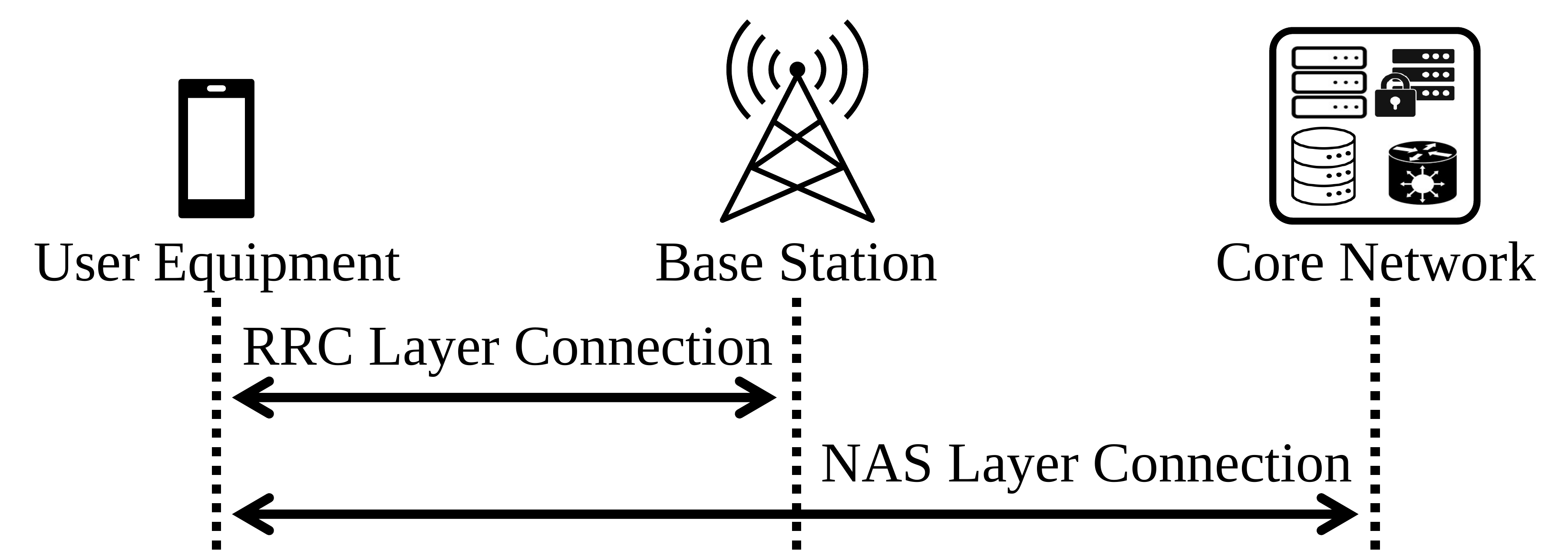}
      \caption{Logical connection including user, base station, and core network.}
      \label{fig:logical_arch}
    \end{subfigure}
    \caption{Cellular networking overview.} 
    \label{fig:overall_architecture}
\end{figure*}

\section{Related Work and Our Approach} 
\label{sec:related_works}
\subsection{Current Practice}
In the \emph{current practice} of 5G, there is no authentication mechanism for the user equipment to authenticate or verify the integrity of the control communications.  
In fact, the current 5G networking implementation and practice does not involve or utilize the base station's public key, which enables cryptographic mechanisms, although there have been recent proposals in research~\cite{singla2021look,hussain2019insecure,ross2024fixing,dong2025securing} and standardization/development~\cite{3gppfbs} to introduce and use the base station's public key. 
%
In contrast to the current practice, our scheme builds the base station's credentials (binding the ID and location to the public key) and authenticates the base station. 

\vspace{-4px}
\subsection{State of the Art Research (SOTA)}
\label{subsec:sta_r}

We compare our scheme with the state-of-the-art research (SOTA), for which we combine and synthesize the research works, making contributions in the certificate construction and use in base station communication. 
Recent research~\cite{hussain2019insecure,singla2021look,gao2021evaluating, ross2024fixing, dong2025securing} and older research~\cite{yi1998optimized,zheng1996authentication} 
proposed using the RRC bootstrapping communications to share the base station's public key and authenticate the base station via digital signature, 
while \cite{lotto2023baron} also authenticated the base station during handover 
using authentication tokens. 
However, the recent research in~\cite{hussain2019insecure,singla2021look, ross2024fixing, dong2025securing} are particularly relevant and similar to our work in that they provide systems and implementation-based study, build on 4G/5G networking, use the SIB1 RRC message, 
and build on the standard cryptographic techniques based on digital signature and public-key-infrastructure (PKI)-based authentication.

For certificate/key delivery, Hussain et al.~\cite{hussain2019insecure} and Ross et al.~\cite{ross2024fixing} use the preinstalled public key of the core network in the Universal Subscriber Identity Module (USIM) 
as the initial anchoring point to generate the certificate chain, and the mobile user verifies the digital signatures of the intermediary nodes (e.g., MME/AMF) and base station through the certificate chain, i.e., using the public keys of those along the chain. 
Building on such key delivery, Hussain et al.~\cite{hussain2019insecure}, Singla et al.~\cite{singla2021look}, Ross et al.~\cite{ross2024fixing}, and Dong et al.~\cite{dong2025securing} digitally sign the base station's radio control communications. In these schemes, the user equipment verifies three digital signatures (core network, MME/AMF, SIB1) to authenticate the base station control communications that induce computational load at the user equipment (e.g., foregoing certificate chaining~\cite{hussain2019insecure, ross2024fixing} or multi-party computation~\cite{singla2021look, dong2025securing}). 

\vspace{-4px}
\subsection{Our Contributions Beyond SOTA}

We compare our work with 
SOTA based on the synthesis of the collection of research works described in Section~\ref{subsec:sta_r}. 
Our work partly builds on these SOTA works in that we also rely on the core network's public key delivery via USIM and the digital signature encoding in the RRC SIB1. 
However, our work enables greater security and efficiency than SOTA, which itself outperforms the individual research works. 

Our scheme advances beyond SOTA by taking a systems approach and co-designing the certificate construction, delivery (involving blockchain-based offline delivery), and the multi-factor authentication. 
The synergistic co-design of these processes enables security and performance advantages over SOTA in four critical ways.
%
First, our scheme reduces the number of signature verifications by directly sharing the base station's public key, while the previous research requires three digital signature verifications (for the core network, MME/AMF, and SIB1/base station). 
Second, the base station certificate includes the base station's geographical location, enabling the user equipment to check and verify the location authenticity of the base station. The location verification detects the wormhole threats while SOTA remains vulnerable. 
Third, our scheme delivers the base station's public key, ID (to identify which IDs are certified by the core network), and the location offline via blockchain, while the previous research delivers them along with SIB1 online, i.e., appends the public key in SIB1. Our scheme yields the following advantages: provides a more robust delivery of such offline control communications, enables ECDSA digital signature algorithms of greater security and key length, and reduces the computation and battery power consumption at the user equipment. 
Fourth, the multi-factor authentication uses multiple orthogonal channels to verify the factors presented by the subject base station: the ledger serving as the registered base station list, the real-time location sensing, and the cryptographic signature verification. An attacker needs to compromise all of these channels to break the base station authentication and to inject control communications to get accepted, reducing the threat feasibility. 
We empirically validate these design goals and advantages. 

\subsection{Other Related Work in Malicious Base Station}
\label{sec:related_work}

We describe SOTA in base station public key and authentication in Section~\ref{subsec:sta_r}; we later compare our scheme with SOTA. 
Section~\ref{sec:threat_analyses} also provides a more in-depth analysis of related research and defense than this section after describing the relevant background in Section~\ref{sec:background}.

\subsubsection{Malicious Base Station Threats}
Because there is no base station authentication in current mobile networking, including the most recent 5G, 
a rogue and malicious base station can have the mobile user connect to itself by advertising higher channel quality and signal strength~\cite{yang2019hiding,DBLP:conf/ndss/ShaikSBAN16,hussain2018lteinspector}. The advancements and developments in software-defined radios (SDR) and software-based radio/mobile control also increase the feasibility of malicious base stations, e.g., there are SDR-based tutorials for such implementations~\cite{sdrhacking1,sdrhacking2,sdrhacking3}. Once a gullible mobile user connects to a malicious base station controlled by an attacker, the attacker can launch DNS redirection attack~\cite{rupprecht2019breaking}, bypass the 5G authentication and key agreement (AKA), and/or de-register the user from the core network~\cite{kim2019touching}, downgrade the user to 2G/3G network~\cite{DBLP:conf/ndss/ShaikSBAN16,shaik2019new,3gppfbs,2023karakoc}, drain the user's battery~\cite{shaik2019new,hussain20195greasoner,Ishtiaq2024}, base station resource depletion~\cite{wen20245g}, launch SMS phishing~\cite{zhang2020lies,wen2023thwarting}, and track the mobile user's location (violating the user privacy, critical in mobile computing)~\cite{DBLP:conf/ndss/ShaikSBAN16,hussain2018lteinspector,hussain2019privacy,hussain20195greasoner}. The base station also has dominant control over the user's connectivity availability and can disrupt it via denial-of-service~\cite{DBLP:conf/ndss/ShaikSBAN16,hussain2018lteinspector,shaik2018impact,hussain20195greasoner}. 

Our work is motivated by these threats discovered from previous research. We authenticate the base station to enable the mobile user to distinguish between a legitimate vs. rogue base station. 
Our work is a preventive defense measure, unlike other defense research for passive detection (e.g.,~\cite{dabrowski2014imsi,DBLP:conf/ndss/Li0WC0JZL0L17,alrashede2019imsi,sou2024,saad_10903974,purification_electronics13173474}) or avoidance-based recovery after the attack (e.g.,~\cite{saad_10903974,purification_electronics13173474}). 

\subsubsection{Malicious Base Station Attacks on Radio Control Communication}
The adversary's choice for launching a rogue base station attack is to use the radio control channel between the base station and the user. An attacker can exploit vulnerabilities in radio control communication for handover scenarios~\cite{bitsikas2021don}. Similarly, it also utilizes system information messages to broadcast fake warning/alert messages in 5 G-NR~\cite {bitsikas2022you} and in 4G LTE~\cite{lee2019your}. Our motivation to authenticate the base station using system information messages comes from their findings that show the attacker initiates attack steps by broadcasting the insecure system information messages. 

\subsubsection{Wormhole Defense}
Although wormhole attacks can apply to any wireless network, to the best of our knowledge, there is no previous defense research against wormhole attacks in cellular networking. 
In the broader wireless networking contexts, including ad hoc networks and sensor networks, the defense approaches can be divided into two categories based on whether spatial or temporal information of the sender is required or not. Hu et. al~\cite{hu2003packet} provide a packet-leash-based defense mechanism for wireless ad hoc networks, where the packet leash is the geographical and temporal information of the packet that limits the distance from the sender and lifespan of the packet. On the other hand, there are research works that do not use geographical location or time synchronization information to thwart wormhole attacks. Such research prevents wormhole attacks by estimating the physical distance using distance-bounding protocol~\cite{vcapkun2003sector}, using directional antennas with an accurate set of neighbors~\cite{hu2004using}, monitoring neighbor node behavior~\cite{choi2008wap}, and a cryptography-based approach~\cite{lazos2005preventing}. The recent approaches for detecting and preventing wormhole attacks at the network layer routing in mobile ad-hoc networks include machine learning-based~\cite{tahboush2021hybrid,abdan2022machine}, enhancing routing protocol~\cite{sankara2020modified}. 
Our threat model includes fake base station signal injection, including wormhole as described in Section~\ref{subsec:threat_analyses}. 
Our work defends against the wormhole threat, while the SOTA, which combines multiple research literature, does not provide defense against wormhole.

\vspace{-1px}
\section{Background}
\label{sec:background}
We describe the 5G New Radio (NR) protocol, offline vs. online communications, and the standard cryptographic techniques of public-key digital signature and certificate. 
We build on these backgrounds to describe our scheme in Section~\ref{sec:our_scheme}, including the notations and variables for the digital signature. 
The online communication for our scheme builds on the existing 5G protocol, as we describe in Section~\ref{subsec:our_scheme_existing_5G} (no additional communication packets), while the offline communications using blockchain are new. 
\subsection{Radio Control in 5G Cellular Network: 5G RRC and SIB}
\label{subsec:background_5G}
5G cellular network comprises of: user equipment, 
base station, and core network. Figure~\ref{fig:physical_arch} illustrates the physical connectivity among these entities, including the intermediate routers between the base station (at the networking edge) and the core network. 
The base station acts as a bridge gateway between the wireless and the wired communications, i.e., makes a physical connection with the user equipment through wireless channels and with the core network using wired networking (including routers and switches) for the user equipment to access the remote data and services beyond the immediate base station. For connectivity, user equipment establishes a Radio Resource Control (RRC) connection with the base station first and then sets up Non-Access Stratum (NAS) layer connection with the core network~\cite{3gpprrc}, as shown in Figure~\ref{fig:logical_arch}. 
Our work focuses on the control communications between the user equipment and base station, i.e., the RRC communication. 

There are numerous base stations to provide coverage across geographical regions, e.g., in the US, as of the end of 2021, there were an estimated 420,000 base stations across the major cellular service providers~\cite{totalbss}. 
Each base station provides connectivity provision in multiple cells (typically three or more~\cite{cellmapp} depending on the radio resources, including the MIMO antennas). A base station can be uniquely identified by the cell IDs, and each cell ID is unique within a cellular service provider network~\cite{cellmapp}.  
We thus use the cell ID to identify the base station. 
A base station can have multiple cell IDs, 
and each cell ID requires a public key in our scheme.

Because a base station serves many user equipment within its coverage cells, the base station publicly broadcasts and advertises its presence and the RRC-initiating messages, which are the System Information (SI) messages of Master Information Block (MIB) and System Information Blocks (SIBs). 
The cell-specific System Information Blocks (SIB1) message includes the system parameters for the user equipment to bootstrap the radio connection setup (e.g., the cell ID, network ID, and channel frequency information) and it is required for the RRC and the following digital communications. During the initial radio connection, the user equipment scans for these broadcast messages and chooses a suitable base station cell to set up the RRC connection based on the signal strength of the SI messages. Because SIB1 is regularly and periodically broadcast, unlike some of the other conditionally transmitted SIBs, we use the SIB1 packet to add base station authentication in our scheme. 

The SIB1 message size is determined by the transport block size (TBS), which gets transmitted by the base station at the initial physical layer signaling using 
Physical Downlink Shared Channel (PDSCH) signaling and Downlink Control Information (DCI). The SIB1 message or TBS cannot exceed 2976 bits (372 Bytes) by 5G NR standard~\cite{3gpprrc}. 
\subsection{Offline vs. Online }
\label{subsec:offline_vs_online}

We define online to be during the communication and connectivity provision when the user equipment is in contact with the base station. 
The \emph{online} operations occur when the user equipment wants to use the cellular service and connectivity, and 
begins establishing the RRC control communication with the base station (in which the user equipment responds to the SIB1 to establish the radio connection). 

The \emph{offline} operations occur before the \emph{online} operations, i.e., before the user equipment connects with the base station to initiate the RRC control communications. The offline operations include the user equipment registration, including the Universal Subscriber Identity Module (USIM) setup and any previous communication sessions that the user equipment went through. USIM includes the core network's public key by having it coded in during the user equipment registration. Using such a core network's public key, the core network and user equipment perform mutual authentication and establish the security parameters through Authentication and Key Agreement (AKA) protocol~\cite{3gppsap}. 

Because we define offline to be any previous communications before the online, when the user equipment is communicating for payload and encounters the base station, the offline costs in computing, networking, and energy are significantly cheaper than the corresponding online costs. The online costs delay the communication connection in real time, while the offline costs are in advance of the communication connection.

\subsection{Digital Signature and Certificate Variable Notations}
\label{subsec:background_public_key}

Public-key cryptography provides the standard and widely used cryptographic techniques for the digital signature (providing sender authenticity and message authenticity) and public-key infrastructure or PKI (providing the digital certificate binding the public key to the subject's identity and serving as the root of trust). 
For digital signature, the subject/prover generates a public-private-key pair ($K$ is the public key while $k$ is the private key) and digitally signs a message $m$ using its private key $k$ to generate the digital signature $S_k(m)$, which involves the public-key cipher algorithm using the private key $k$ on the hash of the message $m$. The verifier receives the message $m$ and the signature $S_k(m)$, which are the two inputs for the verification $V_K(m,S_k(m))$. The verification i) computes the hash of $m$ and ii) applies the public-key cipher algorithm using the public key $K$ on $S_k(m)$. If the two results in the same, then the signature is verified; if not, the verification fails. The verification $V_K(m,S_k(m))$ results in the Boolean. The key inputs for signing $S$ and verification $V$ are in subscripts while the message inputs are in parentheses because the key inputs are fixed while the messages change, i.e., key refreshes are significantly less frequent than the message changes. While any entity involved in networking can generate the public-private key pair and digitally sign, we focus on the base station as our goal is to authenticate the base station using its public key, i.e., $K$ and $k$, respectively, are the base station's public key and private key in our work.

\vspace{-1px}
\section{Threat, Vulnerability, and Defense Analyses}
\label{sec:threat_analyses}


\renewcommand\tabularxcolumn[1]{>{\Centering}m{#1}}
\newcolumntype{Y}{>{\Centering\arraybackslash}X}
\newcolumntype{C}{>{\Centering\arraybackslash\hsize=.2\hsize}c}

\begin{table*}[t]
\begin{center}
\caption{
The comprehensive threat analysis in base station authentication and 5G RRC control communication. \\ 
The threat vectors generally increase in sophistication from top to bottom within each threat type. The right half of the table provides defense analyses in current practice (CP), state-of-the-art (SOTA), and Our Scheme (OS). SOTA is a collection of research papers and efforts as shown in the Ref. column.}
\label{table:threat_analysis}
\begin{adjustbox}{width=\textwidth}
\begin{threeparttable}
\begin{tabularx}{\textwidth}{|Y|Y|Y|C|YV{2.5}V{2.5}C||C|C||C|C|C|}
\boldhline
\textbf{Threat type} & \textbf{Vulnerability} & \textbf{Threat vector} & \textbf{Ref.} & \textbf{Threat impact} & \textbf{CP} & \textbf{SOTA} & \textbf{Ref.} & \textbf{OS} &  \textbf{Sec. Para.} &  \textbf{\S} \\ \boldhline

\multirow{3}{1.5cm}{New Base Station Threat} 

& Known control protocol messages & Transmit fabricated base station ID and RRC message & ~\cite{rupprecht2018security} & Authentication failure and connect to the attacker & {\Large $\times$} & {\Large $\checkmark$} & ~\cite{hussain2019insecure,singla2021look,ross2024fixing,dong2025securing} &  {\Large $\checkmark$} & $ID$ & \S\ref{subsubsec:ID}  \\ \cline{2-11} 

 & No public-key certification & Distribute fabricated public key & ~\cite{3gppfbs} & Authentication failure and connect to the attacker & {\Large $\times$} & {\Large $\checkmark$} & ~\cite{hussain2019insecure,singla2021look,ross2024fixing} & {\Large $\checkmark$} & $ID$ & \S\ref{subsubsec:ID}   \\ \hline

\multirow{3}{1.5cm}{Base Station Spoofing} 
& Insecure system information messages & Transmit fabricated base station ID & ~\cite{rupprecht2018security} & Force the user equipment to connect with the fake base station & {\Large $\times$} &{\Large $\checkmark$} & ~\cite{hussain2019insecure,singla2021look,ross2024fixing} & {\Large $\checkmark$} & $K$ & \S\ref{subsubsec:SIB1}  \\ \cline{2-11} 

 & Insecure broadcast messages and measurement report & Handover exploitation & ~\cite{bitsikas2022you} & Denial of service, user equipment battery drain, cell outage & {\Large $\times$}  &  {\Large $\triangle$} & ~\cite{hussain2019insecure,singla2021look,lotto2023baron,ross2024fixing} & {\Large $\triangle$} & $K$  & \S\ref{subsubsec:SIB1} \\ \cline{2-11} 
 
 & Insecure broadcast and paging messages & Dispatch false public warning/alert messages & ~\cite{bitsikas2021don} & Authentication failure and false alarms (e.g. emergency alarms) & {\Large $\times$} & {\Large $\triangle$} & ~\cite{hussain2019insecure,singla2021look,ross2024fixing} & {\Large $\triangle$} & $K$ & \S\ref{subsubsec:SIB1} \\ \cline{2-11} 

   & Insecure radio \& NAS control messages & User equipment identification & ~\cite{2023karakoc,shaik2019new} & User Device capability identification & {\Large $\times$} & {\Large $\triangle$} & ~\cite{hussain2019insecure,singla2021look,ross2024fixing} & {\Large $\triangle$} & $K$ & \S\ref{subsubsec:SIB1} \\ \hline

  
  \multirow{3}{1.5cm}{Control Communication Injection}
  
  & Insecure radio \& NAS control messages & Bidding down attack & ~\cite{2023karakoc,shaik2019new} & Downgrade service from 5G to 4G to 3G to 2G & {\Large $\times$} & {\Large $\triangle$} & ~\cite{hussain2019insecure,singla2021look,ross2024fixing} & {\Large $\triangle$} & $K$ & \S\ref{subsubsec:SIB1}  \\ \cline{2-11} 
  
  & Insecure initial RRC connection procedure messages & Man-in-the-Middle attack & ~\cite{hussain20195greasoner} & User equipment in limited service/security context & {\Large $\times$} & {\Large $\triangle$} & ~\cite{hussain2019insecure,singla2021look,ross2024fixing} & {\Large $\triangle$}  & $K, t$ & \S\ref{subsubsec:SIB1},\S\ref{subsubsec:t}  \\ \cline{2-11} 
  
  & Lack of temporal and spatial information & Wormhole attack & ~\cite{hu2003packet}\tnote{*} & Bypass the cryptographic authentication & {\Large $\times$} & {\Large $\times$} & $-$ & {\Large $\checkmark$} & $L, t$  & \S\ref{subsubsec:L},\S\ref{subsubsec:t} \\ \boldhline
\end{tabularx}
 \begin{tablenotes}
\item [*] This research is not specific to telecommunications and base station. \hfill {\small[Degree of defense - Complete ($\checkmark$), Partial ($\triangle$), None ($\times$)]}
 \end{tablenotes}

\end{threeparttable}
\end{adjustbox}
\end{center}
\end{table*}

We take a systems approach for our research as it spans across multiple processes in the telecommunications system, including both offline and online. To highlight such a systems approach and identify how and against which threats our scheme makes security contributions, we describe our threat model and then conduct comprehensive threat and defense analyses on the base station authentication and radio control communication message integrity. 
%
Our analyses reflect the ongoing research. To align with our research scope, we focus on the threats of 5G and the RRC communications between user equipment and the base station. The related research in broader 5G communications security is discussed in Section~\ref{sec:related_work}, including those based on previous-generation cellular technologies (e.g., 4G) and those based on security control with the core network (e.g., NAS). 
We also describe the defense research relevant to these threats and compare our scheme contributions with them. 
For the comparison with our scheme vs. the SOTA, we collect and aggregate multiple research works, and our scheme advances the security even further if compared with the works individually. 
Table~\ref{table:threat_analysis} summarizes our threat and defense analyses.

\subsection{Threat Model}
\label{subsec:threat_model}
We consider an attacker acting as a malicious and rogue base station. The attacker is equipped with radio hardware, well-versed in telecommunications protocol, and within the transmission range of the victim user equipment. 
The attacker can inject communications and modify or generate the relevant control communication fields, such as ID, public key, or location. 
The attacker can also listen and capture the communications between the legitimate base station and the user equipment to use that to enable more sophisticated threats. For example, as one of the more advanced threats, we consider a wormhole threat where an attacker uses the captured information from one cell location to craft the communications at another cell location. 

Such a threat model based on a radio-equipped attacker is of high risk (high threat feasibility) because it does not require the system compromise of the victim user equipment or the digital credential compromise of the telecommunications service provider infrastructure, such as the core network's public key. 
For example, there are public SDR tutorials to build most of the aforementioned attacker capabilities, e.g.,~\cite{sdrhacking1,sdrhacking2,sdrhacking4,sdrhacking3}. 
Our threat is of high risk also because the base station's control communications are inherently broadcasting (public advertisements) and the current practice has no key or authentication protections. 
Our threat model aligns well and is comparable with that used by other previous research works, described in greater detail in \sour{Section~\ref{subsec:sta_r}}.


\vspace{-6px}
\subsection{Threat, Vulnerability, and Impact Analyses}
\label{subsec:threat_analyses}
We divide the threats in 5G and the RRC communication that relate to our research into three types: new base station threat, base station spoofing, and control communication injection. We analyze vulnerabilities, threat vectors, and related threat impact for each type of threat which are identified by previous research. We present the threats in the order of increasing sophistication and impact. 
As illustrated in Table~\ref{table:threat_analysis}, insecure control messages (i.e. control messages without authenticity and integrity protection) in the RRC communication are the prime vulnerabilities in the 5G-NR cellular network that the attacker exploits. The attacker takes vulnerabilities in RRC communication as an entry point for luring the benign user equipment to get connected to it and launch attacks beyond the radio control layer. 

An attacker with extensive knowledge of cellular networks and possessing software-defined radios can install a malicious new base station and start informing nearby user equipment about its presence with higher signal strength by broadcasting fabricated system information and RRC messages~\cite{rupprecht2018security}. The lack of public key certification~\cite{3gppfbs} of the base station prevents user equipment from identifying the legitimate base station from the malicious one. As a consequence, the user equipment establishes a radio connection with the malicious base station.

\begin{figure}
    \centering
    \begin{subfigure}{0.22\columnwidth}
      \centering
      \includegraphics[width=\textwidth]{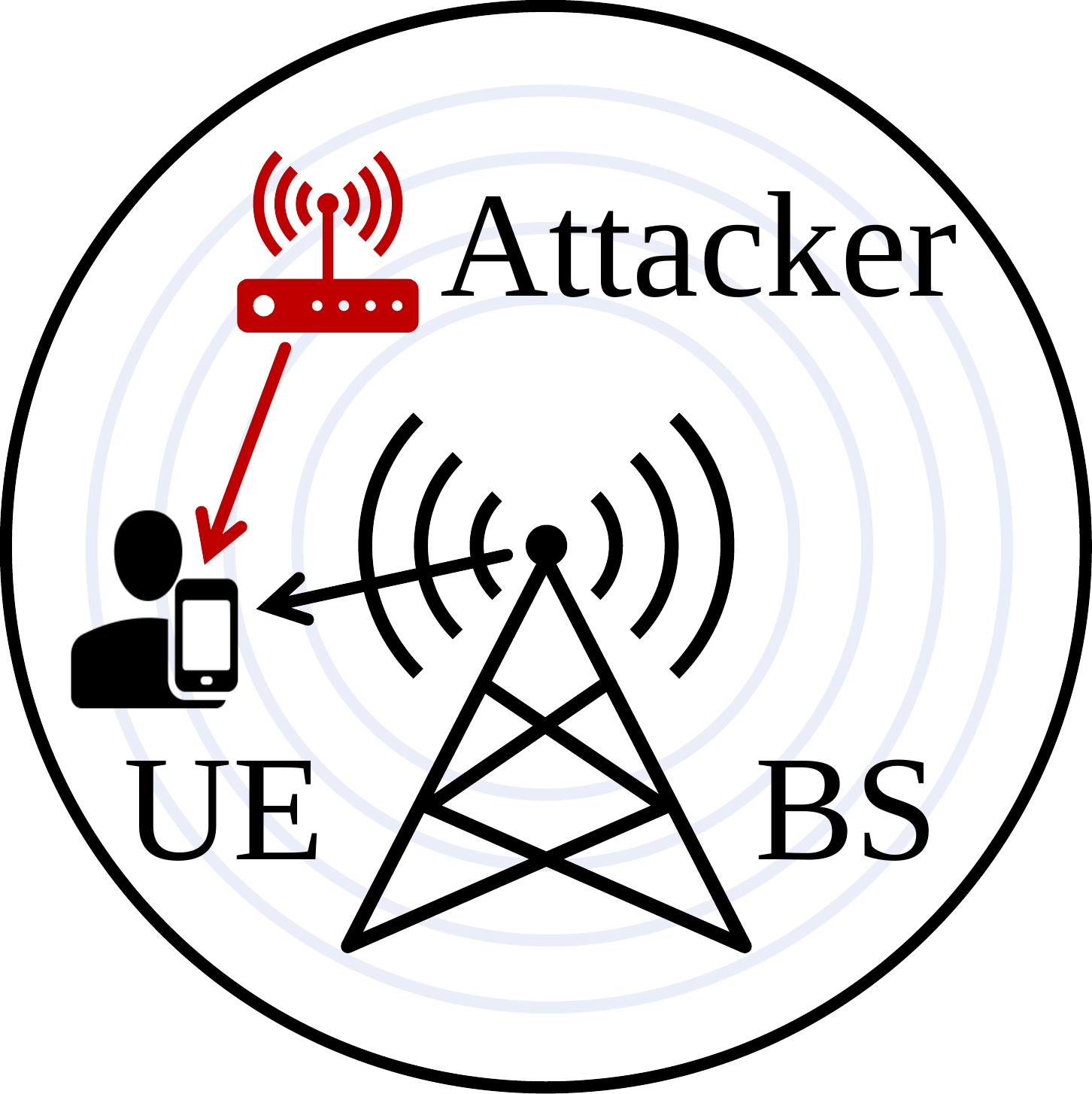}
      \caption{Spoofing}
      \label{fig:attacK_impersonate}
    \end{subfigure}
    \quad
    \begin{subfigure} {0.22\columnwidth}
      \centering
      \includegraphics[width=\textwidth]{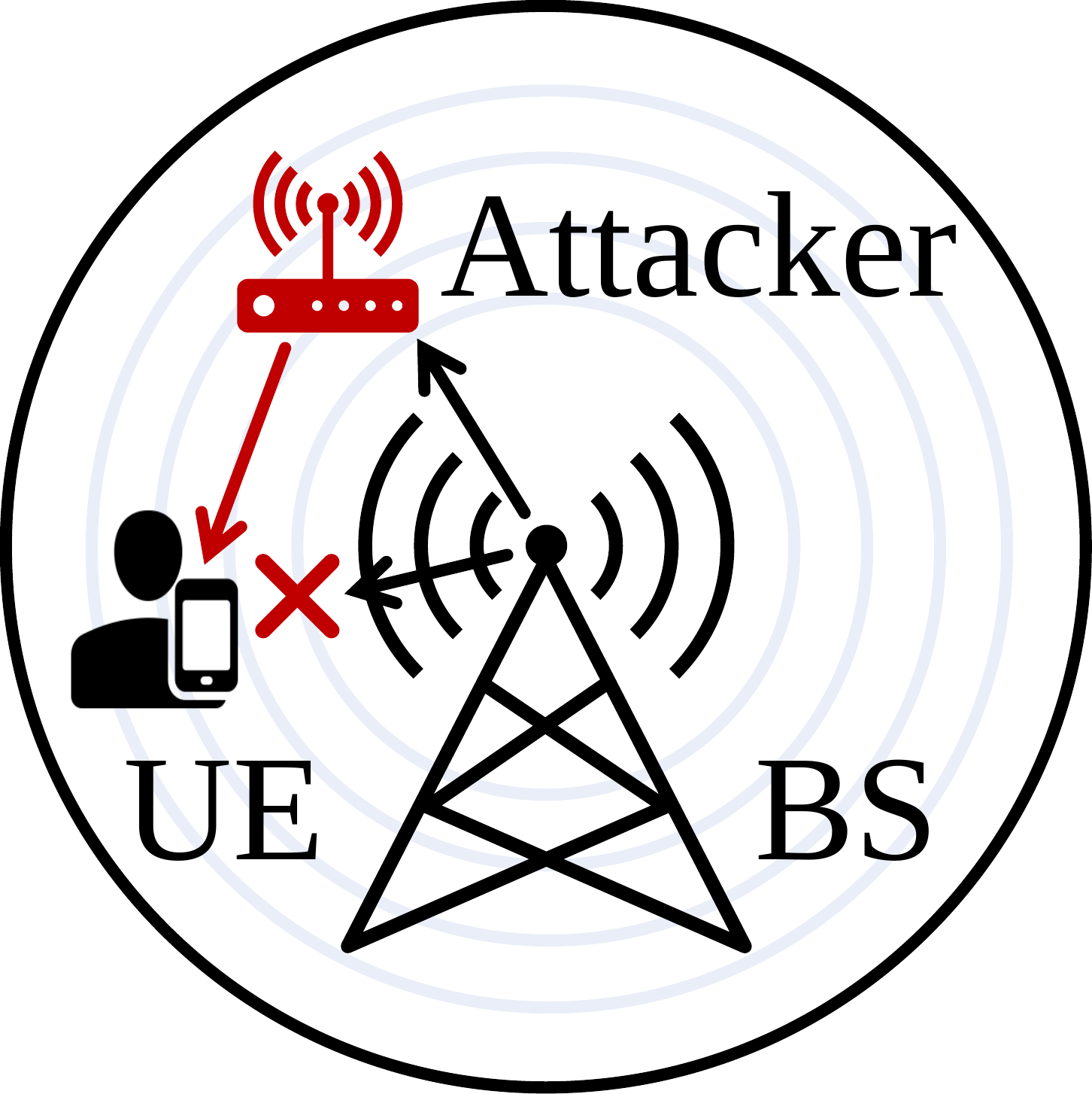}
      \caption{MITM} 
      \label{fig:attacK_replay}
    \end{subfigure}
    \quad
    \begin{subfigure} {0.45\columnwidth}
      \centering
      \includegraphics[width=\textwidth]{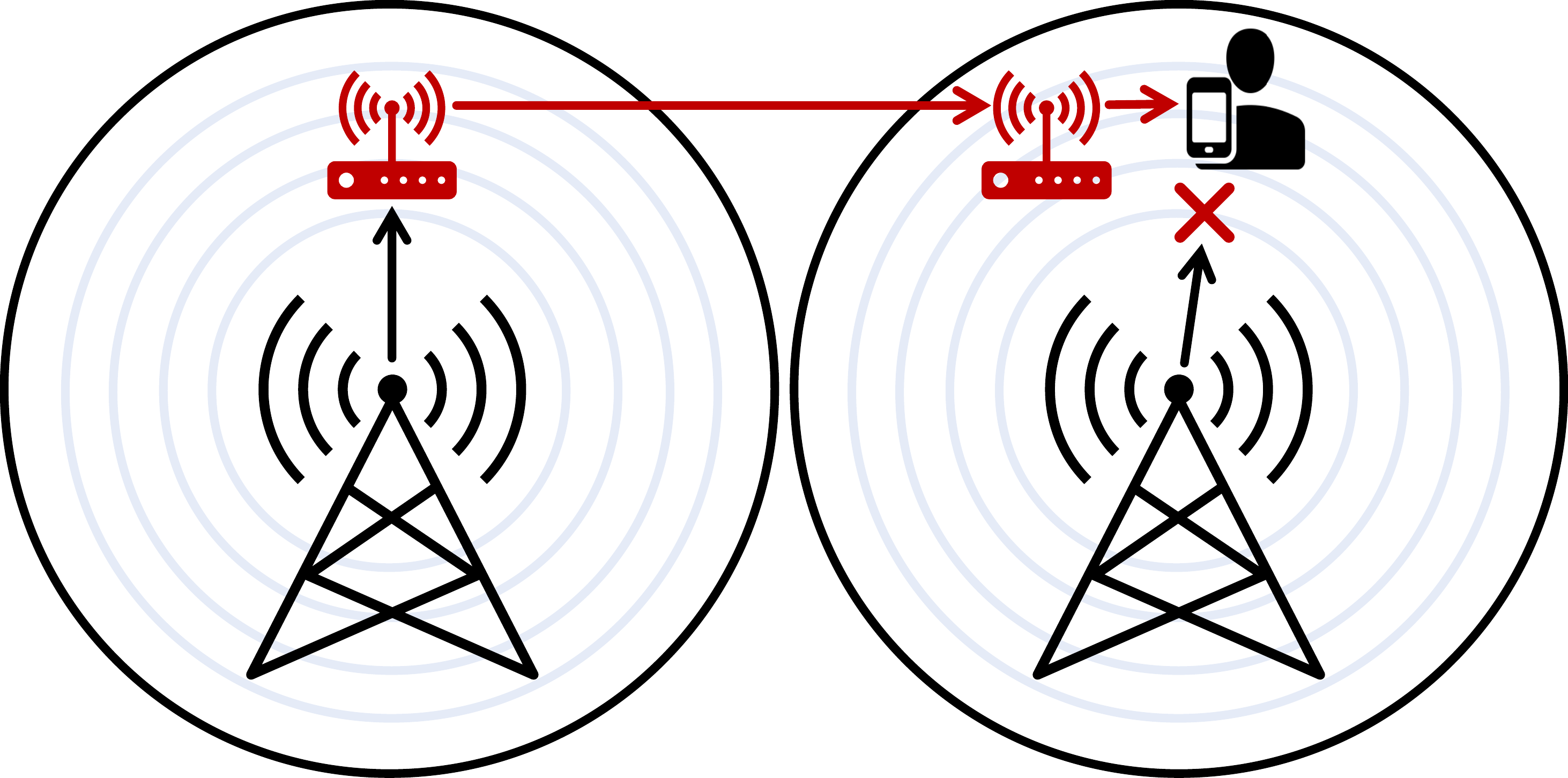}
      \caption{Wormhole}
      \label{fig:attacK_wormhole}
    \end{subfigure}
    \caption{Our threat scenario consists of spoofing attack, man-in-the-middle (MITM) attack, and wormhole attack. \vspace{-15px}}
    \label{fig:attack_scenario}
\end{figure}

The next advanced threat in RRC is base station spoofing, where an attacker impersonates a legitimate base station by capturing its system information messages and broadcasting those later on with higher signal strength to force the user equipment to get connected to it. Figure~\ref{fig:attacK_impersonate} illustrates this attack scenario. The broadcast system information messages have no authenticity or integrity protection, hence the user equipment impulsively trusts the messages and initiates RRC connection with the malicious base station~\cite{rupprecht2018security}. Once the user equipment gets connected to the malicious base station, it can exploit handover using insecure measurement reports. The user equipment collects signal strength measurements and generates measurement reports based on system information messages of nearby base stations, and the report itself is not authenticity or integrity-protected~\cite{bitsikas2021don}. Similarly, the attacker can send public warning system-based paging messages along with the system information messages (SIB6$\sim$SIB8) to dispatch false alarms to user equipment~\cite{bitsikas2022you}. The detrimental impacts of these threat vectors on user equipment are denial of service, battery power drain, and receiving false alarms. Furthermore, the attacker can reveal user equipment device capabilities by exploiting vulnerabilities in NAS layer communication by base station spoofing~\cite{2023karakoc,shaik2019new}. 

The control communication injection threats can have even more harmful impacts. The attacker can inject control communication by acting as a man-in-the-middle (shown in Figure~\ref{fig:attacK_replay}) to cause service downgrade from 5G to 4G to even lower generation networks~\cite{2023karakoc,shaik2019new}, which is known as a bidding down attack. Consequently, the user equipment may operate in lower or no security mode with limited service accessibility, which results in user traffic monitoring, DNS redirection, location tracking, etc. The vulnerabilities of insecure initial RRC connection procedure and NAS control enable the attacker to launch such attacks by injecting or modifying the contents of the messages~\cite{hussain20195greasoner}. However, the most sophisticated attack that an attacker can launch is a wormhole attack~\cite{hu2003packet}. The attacker can launch a \emph{wormhole} attack by listening to system information messages from \sour{a legitimate base station} at one location 
and transmitting
them to another location using a malicious base station as shown in Figure~\ref{fig:attacK_wormhole}, where the transmission is almost instantaneous, i.e., enough to thwart timing- and freshness-based mechanisms. 
Wormhole threat can lead to
launch other attacks such as protocol downgrade, malicious warning messages, etc. There is no previous research about wormhole threat in telecommunication, although wormhole attack has been studied in other contexts, including wireless sensor network, IoT network, vehicular ad hoc network, etc., as described in Section~\ref{sec:related_work}. 

\vspace{-2px}
\subsection{Defense Practice and Research Analyses}
\label{subsec:defense_research}

In defense of the threats imposed by malicious base stations described in Section~\ref{subsec:threat_analyses}, we present the defense analyses in this section. As shown in Table~\ref{table:threat_analysis}, we present the degree of defense against each threat vector in the current practice (CP), state-of-the-art (SOTA), and our scheme (OS). We define SOTA \sour{in Section~\ref{subsec:sta_r}} as a collection of research papers and efforts (multiple research papers) to prevent malicious base station attacks, i.e., we aggregate the research papers to address different functionalities comparable to our scheme. 
We define the degree of defense as complete ($\checkmark$) when the scheme mitigates all the vulnerabilities of a threat vector, as partial ($\triangle$) when the scheme provides defense against a threat vector but fails to address one or more vulnerabilities, and as no defense ($\times$) when the attacker can exploit all the vulnerability of a threat vector. The current practice provides no defense ($\times$) against the threat vectors. 

SOTA authenticates the base station to distinguish the malicious base stations building on~\cite{hussain2019insecure,singla2021look,lotto2023baron,ross2024fixing}. 
\cite{hussain2019insecure,singla2021look,ross2024fixing} presents digital signature-based base station authentication using the system information messages to prevent the user equipment from getting connected to the malicious base stations during initial bootstrapping. ~\cite{hussain2019insecure} utilizes \sour{first two system information messages i.e. SIB1 and SIB2} in 4G-LTE to authenticate the base station whereas ~\cite{singla2021look} uses \sour{SIB1 only} in 5G-NR. Both research append the cryptographic keys on the \sour{system information} message, which are verified by the user equipment using the core network public key preinstalled into the USIM card. With respect to public key delivery and securing the broadcast system information, they provide a complete defense ($\checkmark$) by addressing the vulnerabilities. However,~\cite{hussain2019insecure,singla2021look,ross2024fixing} do not provide authenticity of insecure RRC measurement reports, paging messages, and other RRC communications which implies they provide a partial defense ($\triangle$) for the threat vectors of handover exploitation, dispatching false alarms, user equipment identification, bidding down attacks, and man-in-the-middle attacks. On the other hand, ~\cite{lotto2023baron} authenticates the base station during only the handover process by providing authenticity of the measurement reports using authentication tokens. The vulnerabilities of the insecure broadcast message are not addressed in this research, thus making the degree of defense of the scheme partial ($\triangle$) for the handover exploitation threat vector.
SOTA uses only cryptographic digital signature-based authentication to validate a base station. However, the attacker can still launch a wormhole attack by exploiting the lack of temporal and spatial information on the message while \sour{validating} the cryptographic authentication. Hence, SOTA has no defense ($\times$) against it. 


We advance in defending against the threat vectors because our scheme uses information from base station certificates and online communication for multi-factor authentication: $ID$ verification, $L$ verification, $t$ verification, and SIB1 message verification. $ID$ verification using base station $ID$ to prevent new base station threats ($\checkmark$) described in Section~\ref{subsubsec:ID}. 
$L$ verification validates base station location $L$ and $t$ verification validates SIB1 generation timestamp $t$ at the user equipment to prevent wormhole attacks ($\checkmark$) which are described in Section~\ref{subsubsec:L} and Section~\ref{subsubsec:t} respectively. SIB1 authentication using base station public key $K$ provides message authenticity and integrity to broadcast system information message, thus providing partial defense ($\triangle$) which is described in Section~\ref{subsubsec:SIB1}. 

\vspace{-5px}

\begin{figure}[t!]
    \centering
    \includegraphics[width=0.80\columnwidth]{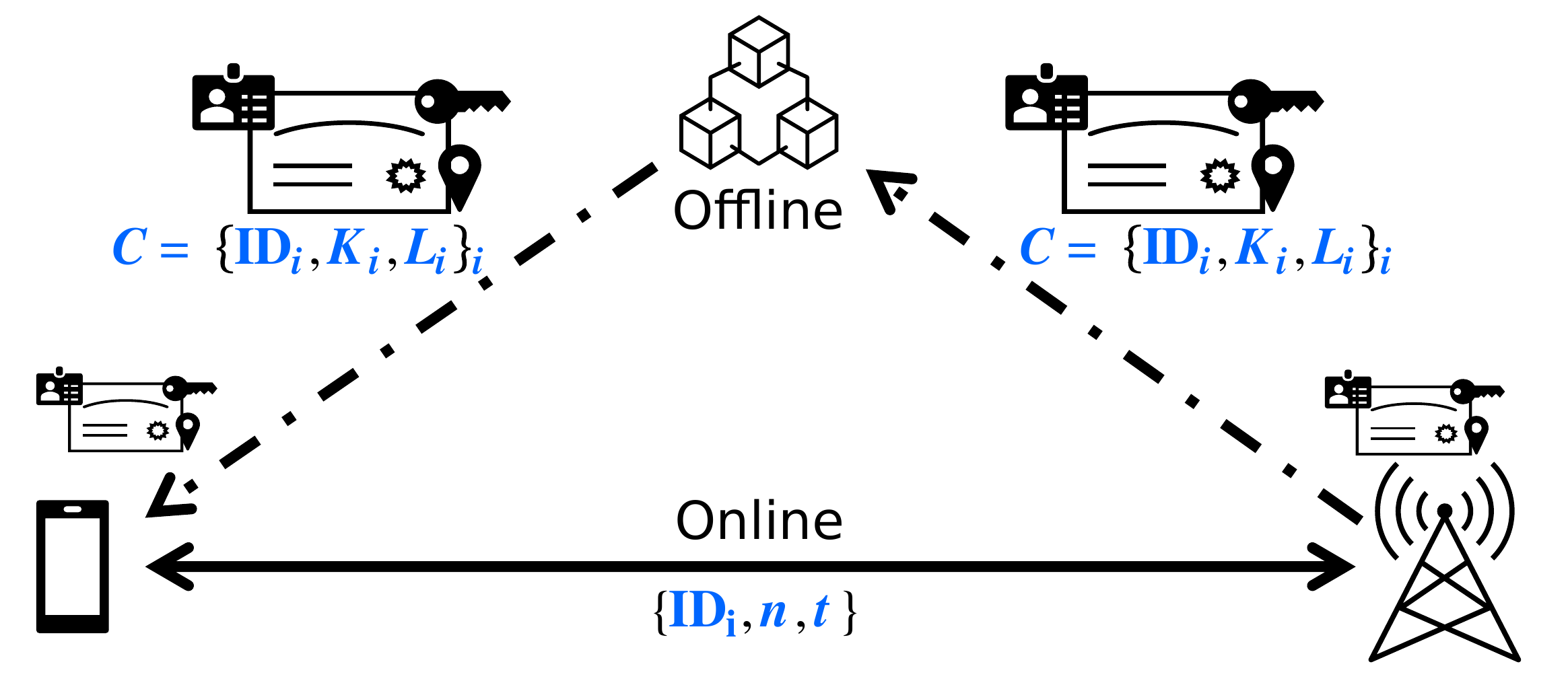}
    \caption{Information delivered during offline (dashed-dotted) vs. online (solid) for our scheme. The certificate database $\mathcal{C}$, including the base station public key, $k$, and its static location $L$ is shared offline, while online involves $ID,n,t$ communication.} 
    \label{fig:key_components}
\end{figure}


\section{Our scheme}
\label{sec:our_scheme}

We design our scheme from the base station digital certificate construction to delivery to its use for multi-factor authentication, as illustrated in Figure~\ref{fig:arrow_diagram}. In our scheme, the construction is offline; the delivery is offline (while the current SOTA research is online as described in Section~\ref{subsec:sta_r}); and the certificate used for multi-factor authentication is online. 

\subsection{Our Scheme Overview in Offline and Online}

\begin{figure}
    \centering
\includegraphics[width=\columnwidth]{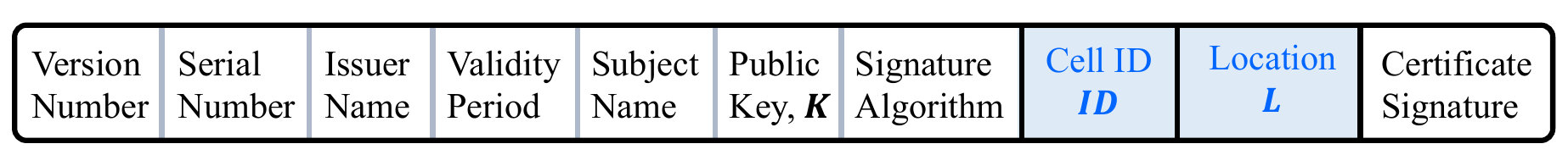}
    \caption{Our base station certificate information fields with the new additions are highlighted in blue. This corresponds to the individual record of a registered base station in the base station certificate database $\mathcal{C}$. 
    \vspace{-2px}
    } 
    \label{fig:cert_contents}
\end{figure}

Our base station authentication scheme consists of two phases: offline (base station certificate construction and delivery) and online (multi-factor base station authentication by using the delivered certificate). Figure~\ref{fig:key_components} shows the entities involved along with delivered information in each phase. In the offline certificate construction and delivery phase, the core network validates and signs the public key of base stations and publishes the base station digital certificate via the blockchain ledger. The core network binds both the base station's cell ID ($ID$) and the geographical location ($L$) with the key ($K$) when digitally signing for the base station certificate. A user equipment that wants to connect to the network receives the base station certificates on the blockchain offline before initiating a radio connection with the base station. The offline blockchain communication includes the base station's certificate (including information fields listed in Figure ~\ref{fig:cert_contents}) and get delivered any time before the radio connection with that base station, e.g., in any of the previous communications by the mobile user. 
During the online multi-factor authentication phase, the base station digitally signs the first radio control communication message, i.e., SIB1, and the user equipment receives and verifies the SIB1 message using the ID, location, and public key of the base station that it gets from the offline blockchain. 
The security of our scheme relies on the security of cryptographic primitives.

In offline, we deliver the digital certificate database of the base stations $\mathcal{C} = \{C_i\}_i$, where each record $C_i = \{ID_i, K_i, L_i, ... \}$ is for the registered base station index $i$ and $i$ iterates across the base stations for each row. For example, the first record/row corresponds to the first registered base station $i=1$, i.e., $C_1 = \{ID_1, K_1, L_1, ... \}$. Each certificate $C_i$ can include other fields as indicated (we implement a concrete certificate based on X.509, which fields are shown in Figure~\ref{fig:cert_contents}), but $ID_i, K_i, L_i$ are the most relevant to our scheme. 
In online, we deliver cell ID ($ID_i$), nonce ($n$), and SIB1 generation timestamp ($t$) appended into the SIB1 message of the base station $i$. Figure~\ref{fig:key_components} illustrates the parameter deliveries between the base station and the user equipment during offline and online. 
This information gets used for authenticating the base station $i$, which yields the Boolean $A$, where $A=1$ means that the base station $i$ has been authenticated (true) vs. $A=0$ not authenticated (false). 
We describe and explain such information deliveries in greater detail in the rest of the sections.

\begin{figure}
    \centering
    \includegraphics[width=\columnwidth]{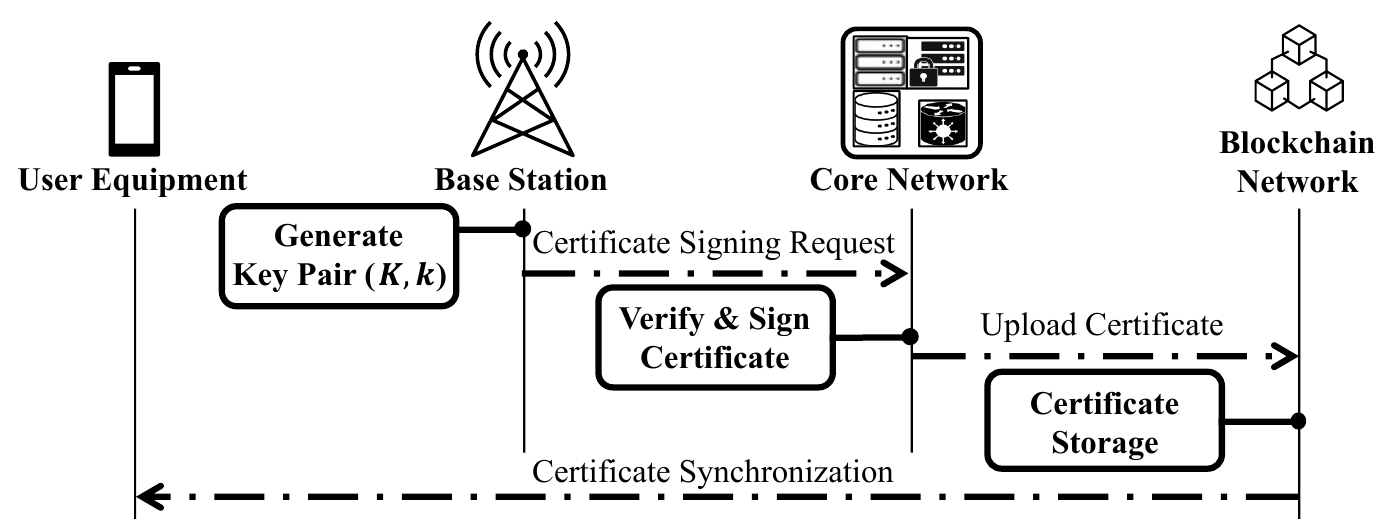}
    \caption{Offline certificate construction and delivery. The certificate construction involves the base station and core network (the inner middle entities), while the delivery further involves the user equipment (left) and blockchain (right).\vspace{-5px}}
    \label{fig:offline}
\end{figure}

\subsection{Base Station Certificate Construction (Offline)}
\label{subsec:our_scheme_cert_construct}

Our base station certificate construction builds on the X.509 standardized certificate, which is popularly used for securing the Internet and web, e.g., transport layer security (TLS) and hypertext transfer protocol secure (HTTPS). However, we adapt its construction specifically for base stations and to enable greater security objectives. 
Figure~\ref{fig:cert_contents} illustrates the data fields within our base station certificate, where our additions beyond X.509 are highlighted in blue. One change we make is the change of formats in the ID from the IP address for Internet/web security to the base station's cell ID ($ID$). 

We add the geographical location of the stationary terrestrial base station in the base station certificate, and the core network uses both the cell ID and the location information to digitally sign and generate the certificate. 
In addition to the cell ID ($ID$) and public key ($K$), our base station certificate includes the location ID ($L$). $L$ is a two-dimensional object containing the longitude and latitude of the base station 
i.e., $L$=$(L_1$,$L_2)$ where $L_1$, $L_2$ are the longitude and latitude coordinate respectively.  
Base station's location information is public information, unlike the user equipment's location and behavior, which require privacy, and is thus appropriate for widespread delivery. For example, in the US, the FCC keeps track of the registered base stations and publishes their cell IDs and the geographical locations~\cite{fcc_cell_data}
which data is also what we use in our prototype implementation later for simulating the real-world base station deployments. 


The certificate construction involves the base station (for key pair generation for the public key and the private key) and the core network (for signing the certificate). 
This section focuses on the certificate construction between the core network and the base station, whose computing and communications are drawn in Figure~\ref{fig:offline} between the middle two entities. While Figure~\ref{fig:offline} illustrates our scheme protocol for both the construction and delivery, Section~\ref{subsec:our_scheme_blockchain} describes the delivery part involving blockchain and others. As shown in Figure~\ref{fig:offline}, the certificate construction process begins with the base station. During the installation of the new base station or when the current digital certificate expires, the base station generates its public-private key pair ($K, k$) for each cell. Then the base station constructs its X.509 certificate (shown in Figure~\ref{fig:cert_contents}) and sends a certificate signing request to the core network. The base station creates separate certificate signing requests for each of its cells and sends them to the core network for signing, generating one certificate per $ID$. 
Upon receiving the signing request, the core network validates the information 
and digitally signs to generate the certificate signature (in the far-right field in Figure~\ref{fig:cert_contents}). 

 \begin{figure}[t!]
    \centering
\includegraphics[width=0.5\textwidth]{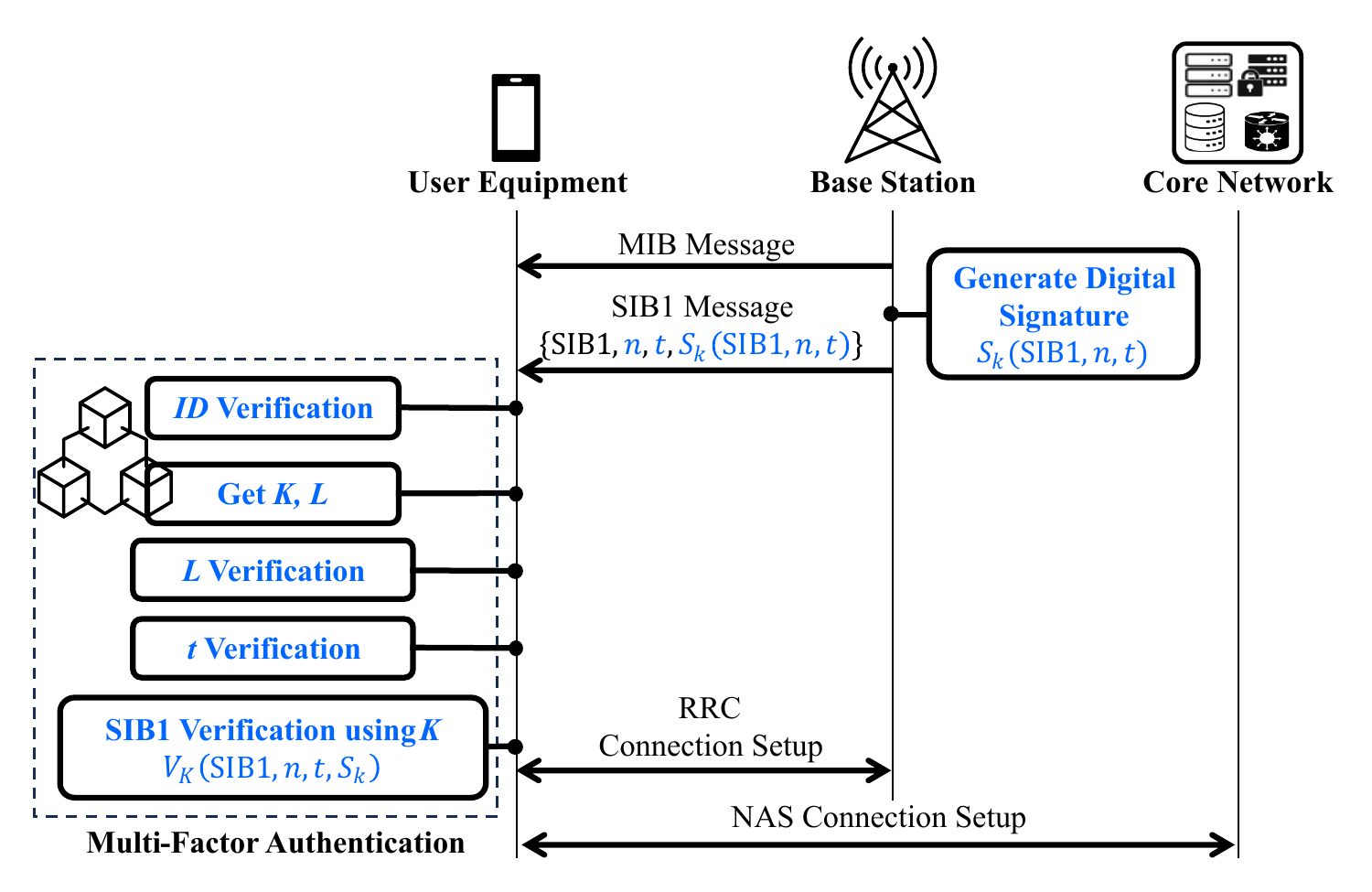}
    \caption{Online multi-factor authentication by user equipment, including the user-side computing for sequential verifications (from top to bottom on the left of the diagram). The blue color indicates our addition in 5G RRC. 
    }
    \label{fig:online}
\end{figure}

\subsection{Certificate Delivery Using Blockchain (Offline)}

\vspace{-2px}
\subsubsection{Blockchain Enabler, Appropriateness, and Effectiveness} 
\label{subsec:our_scheme_blockchain}

Blockchain enables the secure and offline delivery of the base stations' certificates. Blockchain is appropriate for such delivery because, unlike user equipment, the base stations are public entities and publicly advertise their presence via broadcasting SIB1. We store the public information of the base station in the blockchain and avoid the privacy-sensitive or user-equipment-specific information in the blockchain.  
%

Because the service provider infrastructure (core network and base stations) are public entities and do not require anonymity or privacy and because the public keys can be public/known for the security of the public-key digital signature, we use a permissioned but public blockchain where the 
blockchain ledger gets written only by the core network while it can be read/stored/accessed by all the entities in the cellular networking. Therefore, our blockchain forgoes the heavily distributed consensus protocol, e.g., computation-heavy proof-of-work (PoW), to determine which data gets uploaded and by whom. (Technically, our blockchain implementation in Section~\ref{subsec:offline_results} uses PoW consensus but with a difficulty of 0, i.e., all hash function computation for mining is valid.) Also, in contrast to the cryptocurrency blockchain, we do not have the incentivization or competition issues because the service provider infrastructure and its beneficiaries/clients are all aligned in their incentives to provide connectivity and secure cellular connections against the rogue base stations. 
Our use of the blockchain is thus substantially simpler and cheaper than that used in the permissionless applications, such as those for cryptocurrencies and financial transactions. 

Blockchain is effective for our base station certificate delivery. Blockchain provides message authentication integrity anchored by the security-robust cryptographic hash function and public-key digital signature. 
Blockchain 
provides transparent access to the database/ledger thanks to its distributed storage property; in our scheme, the database includes the registered base stations, and blockchain transparency enables accountability. Blockchain also provides automatic synchronization; in our scheme, it can provide the updated base station certificates offline in ad hoc connections, using other non-cellular technology, or from the previous base station connections. 
%
%
We experimentally validate and analyze the appropriateness and effectiveness of blockchain for offline base station certificate delivery in Section~\ref{subsec:offline_results}, while omitting the efficiency validations.
\subsubsection{Certificate Delivery Using Blockchain}
\label{subsec:our_scheme_cert_distribution}
The core network (after responding to the base station's certificate request as described in Section~\ref{subsec:our_scheme_cert_construct}) uploads the base station certificates to the blockchain, as shown in Figure~\ref{fig:offline}. 
The blockchain ledger includes the digital certificates that are signed by the core network and whose subjects are the base stations (as well as the core network itself in the genesis block to enable the writing/uploading of the certificate). 


The blockchain ledger serves as the registered database ($\mathcal{C}$) of registered base stations for the user equipment. The integrity of such records (i.e., only the core network can write and insert) relies on blockchain technology, as described in Section~\ref{subsec:our_scheme_blockchain}. When the base station presents its cell ID ($ID$), if the user equipment does not find it in its locally stored blockchain ledger, then it determines that the base station is not registered and certified by the core network. 

The base station certificate ($\mathcal{C}$) gets delivered offline via blockchain. 
The user equipment uses the core network's public key stored within its USIM card to verify the authenticity of $\mathcal{C}$ and, upon verification, store $\mathcal{C}$ in its local storage. 
This process occurs offline, which is any time before the user equipment becomes active and connects to the cellular base station as described in Section~\ref{subsec:offline_vs_online}. 

\subsection{Online Building on Existing 5G Protocol}
\label{subsec:our_scheme_existing_5G}

Our scheme builds on the existing 5G RRC protocol for compatibility and practicality (facilitating deployment), and this section builds on Section~\ref{subsec:background_5G} but provides greater details for the 5G RRC protocol interactions between the user equipment and the base station. Figure~\ref{fig:online} shows the standard 5G radio control communication protocol messages in black color. The user equipment uses such RRC communication protocol, hence this is a part of the online phase. 
In online, our scheme adds computing and new information networking to the existing 5G protocol. However, the networking only injects new information within the control communication message in SIB1 (i.e., we append the new information on the SIB1 message) but does not introduce additional transmissions beyond that. 

In 5G RRC protocol, the base station broadcasts the regular, periodic MIB and SIB1 messages announcing its presence, including the radio-channel parameters.  
Then the user equipment initiates an RRC connection and sends a unicast communication designed to connect to the base station. 
Then it completes RRC connection setup. 
After a successful RRC connection with the base station, the user equipment proceeds to set up a NAS connection with the core network. We omit the control communication between the base station and the core network in this work as it is out of the scope of our research.

\subsection{Multi-Factor Authentication (Online): Setup, Factors, and Verification Sources}
\label{subsec:MFA_factors}
\vspace{-1px}

In online, when the user equipment connects to a base station, the user equipment implements \emph{multi-factor authentication} to authenticate the base station using multiple factors.
Authentication verification of the subject base station by the user equipment is sequential and uses the multiple factors of $ID$, $L$, $t$, and the digitally signed SIB1 broadcasting message. 
These factors are verified and cross-checked with multiple sources. Section~\ref{subsec:authentication_factors} further provides which information factors are from which source between the offline locally stored ledger, the online communications from the base station via 5G RRC, and the user-equipment sensing of the location $\bar{L}$ and time $\bar{t}$. 

We describe the multi-factor authentication within the 5G RRC communication in Figure~\ref{fig:online} and highlight our scheme in blue font color, whereas the existing RRC protocol is shown in black font color.
Table~\ref{table:threat_analysis} also provides an overview of the mapping between the factors and the threats against which they are effective/defending in their farthest three columns. 

\subsubsection{Signature Generation at Base Station}
\label{subsec:online_signature_generation}

In our scheme, we use the SIB1 for the base station message authentication because it contains the essential control information of the base station and it is periodically broadcast as described in Section~\ref{subsec:background_5G}. As shown in Figure~\ref{fig:online}, 
before broadcasting SIB1 message, the base station generates a digital signature $S_k$(\mbox{SIB1},$n$,$t$) using its private key ($k$). 
We use the SIB1 generation timestamp ($t$) and a nonce ($n$) for replay attack protection during each SIB1 message broadcast. The base station appends the digital signature and other parameters to the \mbox{SIB1} message and broadcasts it to the user equipment. The final SIB1 transmission appends the following: \{\mbox{SIB1},$n$,$t$,$S_k$(\mbox{SIB1},$n$,$t$)\}.

\subsubsection{The Factors and The Verification Sources}
\label{subsec:authentication_factors}
In offline, the user equipment accesses the base station's certificate information $\mathcal{C} = \{ ID_i, K_i, L_i \}_i$, which is stored in its local ledger as described in Section~\ref{subsec:our_scheme_cert_distribution}. 

When the user equipment wants cellular connectivity, i.e., in online, the base station transmits its ID (the cell ID in 3GPP), SIB1, $n$, $t$, and its digital signature, $S_k(\mbox{SIB1},n,t$). 
The user equipment receives these information and uses the base station's public key $K = K_i$ to verify it, where $K_i$ is accessed through the aforementioned offline $\mathcal{C}$. Because we fix the subject base station for our authentication (i.e., we focus on the current base station which the user equipment wants to authenticate and connect to for connectivity if authenticated), we omit the subscript $i$ after accessing it online, including those for $ID_i, K_i, L_i$; if the user equipment switches to another base station, then $i$ changes. 

Furthermore, in online, we sense the location $\bar{L}$ and the time $\bar{t}$ to cross-check with the provided $L$ and $t$, the former from the offline ledger because the base station's location is static. 
The user equipment can control the control parameters of $\tau_d$ and $\tau_t$ for the threshold-based verifications of $L$ and $t$.

\begin{figure}[t!]
    \centering
\includegraphics[width=0.5\textwidth]{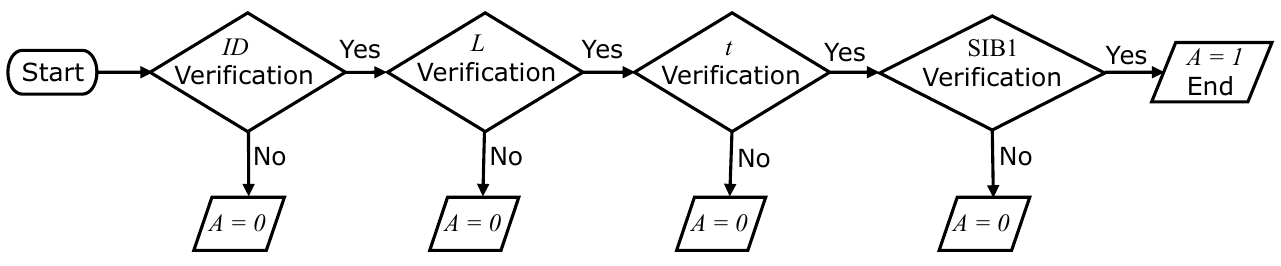}
    \caption{
    The sequential multi-factor authentication by user equipment. For the SIB1 verification, the user equipment verifies the digital signature using the base station's public key $K$.} 
    \label{fig:logic_diagram} 
\end{figure}

\subsection{Multi-Factor Authentication (Online): Sequential Authentication and Verification}
We use the factors and the multiple sources described in Section~\ref{subsec:MFA_factors} for the sequential authentication verifications, depicted in Figure~\ref{fig:logic_diagram}. We describe the step-by-step verification process in this section and explain how each step provides the security protections. While Table~\ref{table:threat_analysis} also provides an overview of which factor/verification provides which security protections against which threats in its farthest three columns, focusing on our scheme, we provide greater details in this section. 

\subsubsection{ID to Verify Base Station Registration}
\label{subsubsec:ID}
The ID verification enables the user equipment to verify whether the base station is legitimately registered by performing a lookup into its locally stored blockchain ledger $\mathcal{C}$. After receiving the SIB1 message from the base station via online communication, the user equipment extracts the $ID$ (which is the Cell ID as described in Section~\ref{subsec:background_5G}) from the message. It performs a ID lookup within $\mathcal{C} = \{ ID_i, K_i, L_i \}_i$. If the $ID$ matches an ID from $\mathcal{C}$, then the ID of the base station is verified. The base station ID verification prevents the new base station threat explained in Table~\ref{table:threat_analysis}. $\mathcal{C}$ contains only the legitimate base stations verified by the core network during offline certificate construction described in Section~\ref{subsec:our_scheme_cert_construct}; therefore, any new base station with a malicious cell ID will not be authenticated. 

The user equipment accesses the location $L$, and the public key $K$ of the base station after ID verification for the location and digital signature verification.

\subsubsection{$L$ for Spatial Integrity}
\label{subsubsec:L}
Upon successful ID verification, the user equipment performs $L$ verification by cross-checking the base station location, $L$, from $\mathcal{C}$, with its sensed location, $\bar{L}$. The user equipment uses a controllable threshold $\tau_d$ for $L$ verification. $L$ verification is successful if the distance between $L$ and $\bar{L}$ is less than $\tau_d$. It limits the validity of the SIB1 message within $\tau_d$, which prevents the wormhole attack because the location verification fails if the attacker transmits the SIB1 message from a distant base station to the user location. More specifically, the $L$ verification disables the attack scenarios illustrated in Figure~\ref{fig:attacK_wormhole}, because the user equipment accepts the SIB1 from the base station within its coverage area controller by $\tau_d$.

\subsubsection{$t$ for Temporal Integrity}
\label{subsubsec:t}
We use $t$ verification as a third factor of authentication, shown in Figure~\ref{fig:logic_diagram}, to ensure the freshness of the online broadcasting control communication of the SIB1 message. The user equipment receives the SIB1 generation timestamp $t$ from online SIB1 communication and compares the timestamp with SIB1 reception time, $\bar{t}$, sensed at the user equipment. Similar to $L$ verification, in that case, the user equipment uses another controllable threshold $\tau_t$ for $t$ verification. If $\bar{t} - t$ is lower than $\tau_t$, then the $t$ verification is successful. In conjunction with $L$ verification, $t$ verification prevents wormhole attacks by limiting the SIB1 message validity period.

\subsubsection{Digital Signature Verification Using $K$ for Message Authenticity}
\label{subsubsec:SIB1}
At the fourth and final verification, the user equipment verifies the SIB1 message by authenticating the digital signature using base station public key $K$, accessed from $\mathcal{C}$ after ID verification. As described in Section~\ref{subsec:authentication_factors}, the user equipment receives the SIB1 message with the digital signature $V_K$($S_k$(\mbox{SIB1},$n$,$t$)) during online SIB1 communication. The user equipment verifies the digital signature $V_K$($S_k$(\mbox{SIB1},$n$,$t$)) to validate the authenticity and integrity of the SIB1 message. It prevents the base station spoofing attack because the digital signature protects the message authenticity and integrity, thus the fake base station cannot modify the radio control parameters to divert the radio connection to itself.

\subsubsection{Authenticated if All Factors Verified} 
The user equipment authenticates the subject base station if all the factors are authenticated and verified. 
A successful multi-factor authentication returns the Boolean $A$, where $A = 1$ corresponds to being authenticated and $A = 0$, an authentication failure. 

As illustrated in Figure~\ref{fig:logic_diagram}, the user equipment performs the chronological computations locally after receiving the SIB1 message from the base station. It verifies the base station ID using $ID$ verification, get the public key ($K$) and location ($L$) from the certificate database ($\mathcal{C}$), verifies base station location using $L$ verification, verifies the SIB1 generation time using $t$ verification, and finally authenticates SIB1 message using $K$. 

\vspace{-2px}
\vspace{-2px}
\section{Implementation and Experimental Setup}
\label{sec:experimental_setup}


\begin{table}[]
\caption{Hardware specifications for the devices used.} 
\centering
\tiny
\label{device_specs}
\resizebox{\linewidth}{!}{
\begin{tabular}{lllll}
\hline
Simulating       & Processor (brand, model \& clock speed)          & Memory\\ \hline
Core Network & Ryzen Threadripper 3960x @ 4.5 GHz   & 64 GB  \\ \hline
Base Station & AMD Ryzen 7 5700U @ 4.3 GHz & 16 GB \\\hline
User Equipment         & Intel 12th N95 @ 3.4 GHz        & 8 GB  \\\hline
3rd Party Node   & Intel Core i7-10750H @ 2.6 GHz       & 16 GB \\ \hline
\end{tabular}
}
\vspace{-2px}
\end{table}

Because we implement both the offline blockchain-based key delivery and its use in the online 5G RRC communications, our implementations include both local (direct communication link) and remote setups. 

\begin{figure}[t!]
\centering
\includegraphics[width=0.48\textwidth]{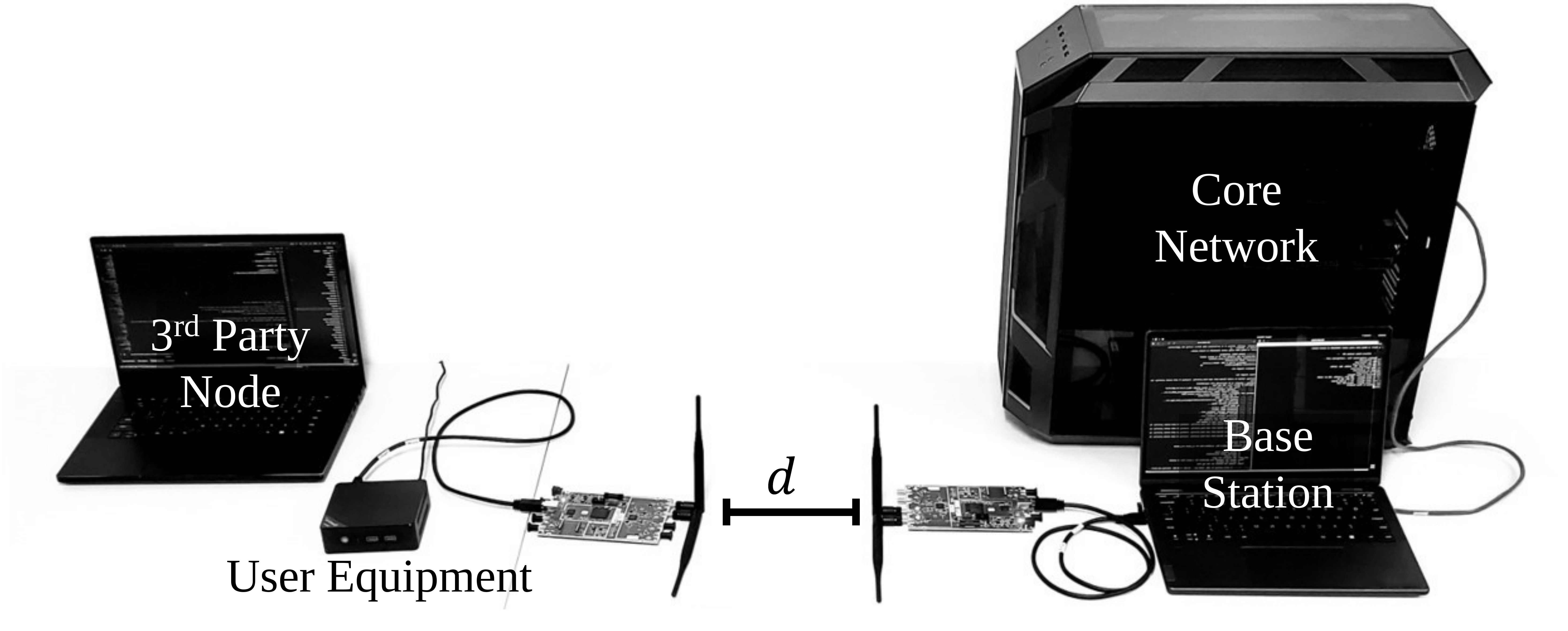}
\vspace{5px}
\caption{Experimental setup with the user equipment, base station, core network, and a 3rd party node. USRP B210 SDR is used for 5G communications. The 3rd party node communicates the blockchain via Wi-Fi or Bluetooth.\vspace{-2px}} 
\label{fig:experimental_setup_image}
\end{figure}

\subsection{Entities} 
We simulate the different nodes involved in 5G RRC and blockchain in our scheme. These nodes are User Equipment, Base Station, Core Network, and 3rd Party Node as depicted in Figure~\ref{fig:experimental_setup_image}. The 3rd Party Node transmits the blockchain communication to the user equipment and represents anybody it communicates through offline, 
for example, ad hoc node or the previous base station. 

For the frontend radio communications between the user equipment and the base station, we use two USRP B210 software-defined radio (SDR) hardware with $d$ distance apart. 
The hardware and processor specifications of the machines that we use for the simulated nodes are in Table~\ref{device_specs}. The highest-rated processor/CPU (AMD Ryzen Threadripper 3960x) is used to simulate the core network, and the second highest-rated processor (AMD Ryzen 7 5700U) simulates the base station. The user equipment hardware (Intel 12th Gen N95 mini PC) is also comparable to that of the 5G smartphone computing capabilities.

\subsection{Offline (Blockchain)} 
We use Go Ethereum v1.10.15 local private blockchain (i.e., independent of the Ethereum cryptocurrency Mainnet) and Solidity v08.1 smart contract for the blockchain implementations. 
We locally store the ledger using Redis 7.0.9 caching to make the real-time online operation more efficient, as Ethereum creates a new transaction for the ledger read, which is unnecessary for our scheme. 
We implement offline blockchain communications using diverse communication channels in both local ad hoc and remote infrastructure-based environments. 
We vary the communication protocols to demonstrate the diverse and robust ways that the blockchain can deliver the base station's public key 
c.f., Bitcoin even gets delivered via unconventional satellite networking~\cite{blockstream}. 
For the ad hoc local network, the 3rd party node and the user equipment connect to Wi-Fi or Bluetooth to distribute the blockchain ledger.  
For the remote Internet-infrastructure-based networking, the remote server (Google Cloud servers in domestic and international locations) connects to our user equipment via TCP/IP for the blockchain/key delivery. 

\subsection{Online (RRC)}
For the online implementation, we use srsRAN v22.10~\cite{srsran}, an open-source 4G/5G prototyping software to incorporate our scheme. We modify the RRC protocol to implement our scheme as shown in Figure \ref{fig:online}. We use OpenSSL v1.1.1 for the digital signature and x.509 certificates, whose cryptographic operations are incorporated in both the online RRC and offline blockchain.

\subsection{Fake Base Station Threat Implementation}
\label{subsec:implementation_threat}
We experiment and analyze the security effectiveness of our scheme against a fake base station attacker described in section~\ref{subsec:threat_model}. 
We implement the fake base station threats described in Section~\ref{subsec:threat_analyses} while focusing on threats related to SIB1 broadcast messages. The fake base station can broadcast an SIB1 message with its fabricated cell ID, spoof a legitimate base station, and launch a wormhole attack.
We collect the operation log files from srsRAN for the attack scenarios to measure the user-equipment effect and response under attack and use these for our analyses. 

\section{Experimental Analyses}
\label{sec:experimental_analyses}

\begin{figure}[t!]
\centering
\includegraphics[width=0.48\columnwidth]{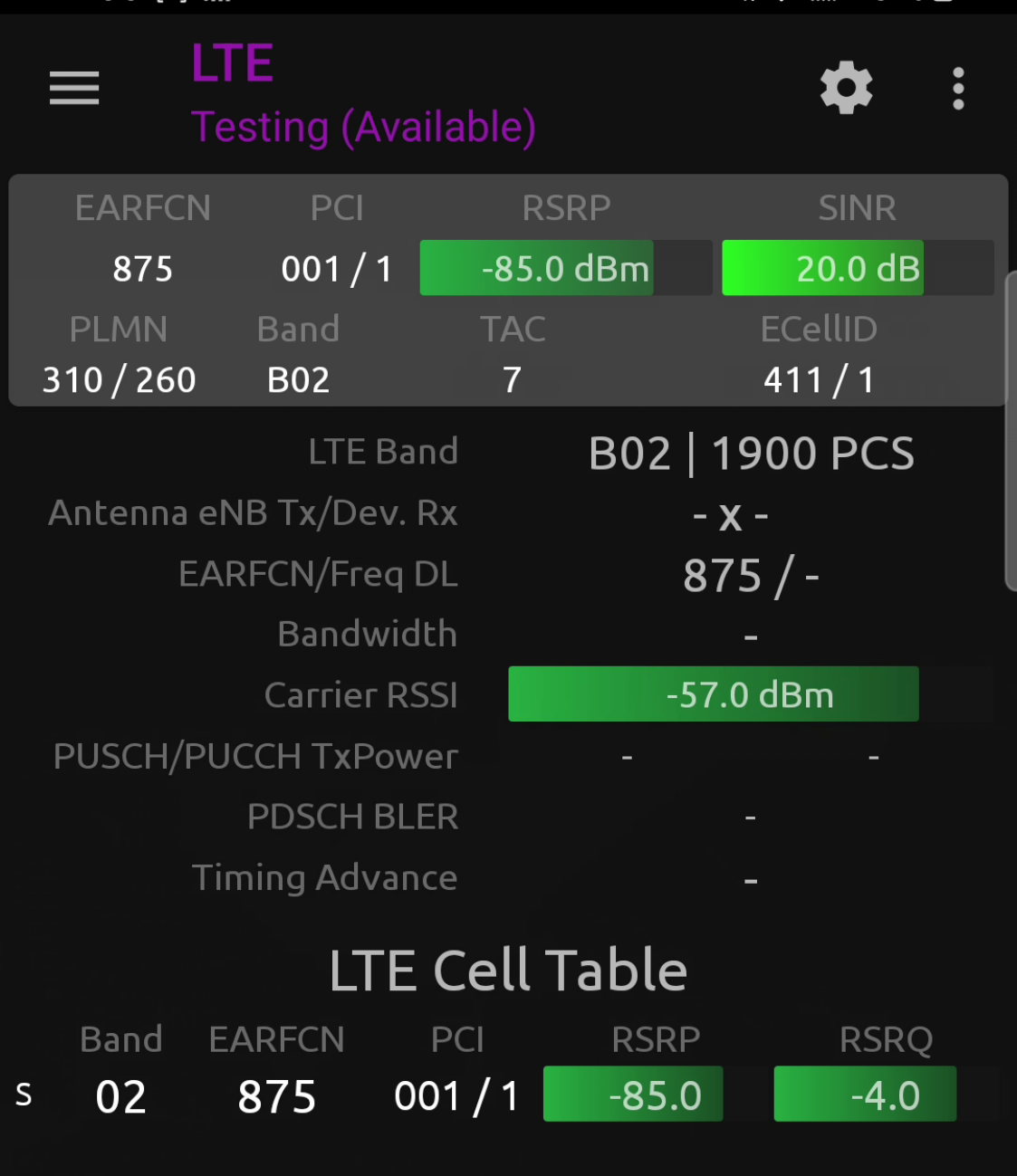}
    \caption{A fake base station takes over connectivity from a legitimate network of Android user equipment in 4G/LTE.}
    \label{fig:fbs_attack}
\end{figure}

\begin{figure}[t!]
\centering
\includegraphics[width=1\columnwidth]{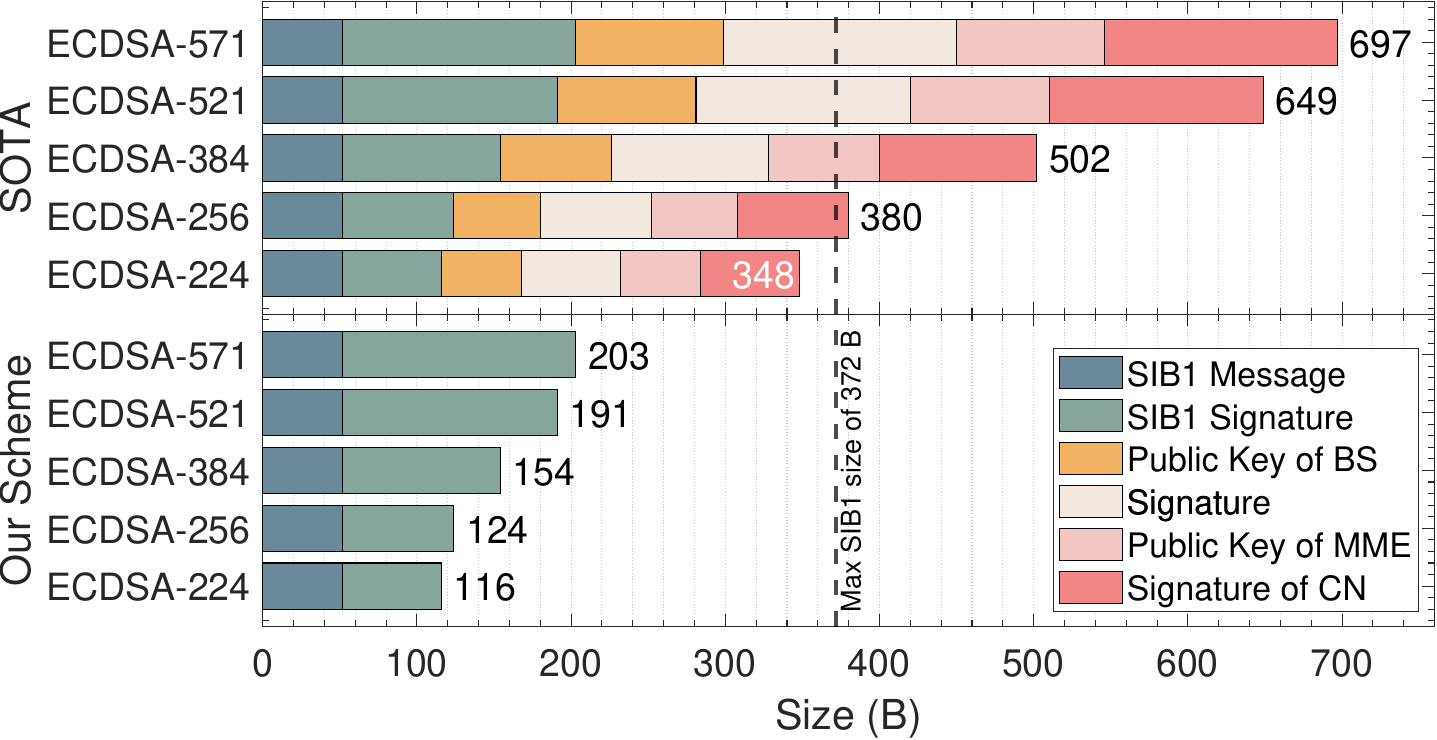}
    \caption{SIB1 packet size comparison between our scheme and SOTA with different signature algorithms. The vertical line shows the SIB1 size limit of 372 Bytes.\vspace{-7px}}
    \label{fig:sib1_packet_size_i}
\end{figure}

\subsection{Security Analyses of the Scheme} 
\label{subsec:security_analysis}

We empirically verify the security effectiveness against the threats of New Base Station, Base Station Spoofing, and Connection Injection. A fake base station can hijack the connectivity without our scheme; it cannot against our scheme. 

\subsubsection{Vulnerable Without Our Scheme}
Without our scheme to authenticate the base station, we empirically validate that the fake base station can take over 4G/5G, as long as it transmits with enough signal power to the victim user equipment to exceed its channel SNR over the other legitimate base stations. 
Figure~\ref{fig:fbs_attack} shows the fake base station taking over an Android user equipment's connectivity to real-world cellular operators; we test against two of the three major operators in the US and show that the fake base station can successfully launch the threat. 


\subsubsection{Enables Greater Cryptographic Strength}
\label{subsubsec:greater_security}
Our scheme, utilizing the offline delivery of the certificate and public key, enables more efficient communications, i.e., fewer bits to transfer in the RRC SIB message. 
As shown in Figure~\ref{fig:sib1_packet_size_i}, our scheme can support all the ECDSA ciphers with varying key lengths and security strengths, while the SIB1 size limit of 372 Bytes (as described in Section~\ref{subsec:background_5G}) prohibits the use of ECDSA with greater keys beyond ECDSA-224 in SOTA. In addition to theoretically analyzing it based on our understanding of the RRC SIB1 protocol and ECDSA, we verified this 
in our prototype implementation.  %
While this section focuses on the increased cryptographic security strength against brute-force attackers, we also experimentally validate and analyze the reduced computation and energy use because of the communication data efficiency in Section~\ref{subsec:results_online}. 

\subsubsection{Against New Base Station Threat} 
Our scheme prevents new base station attacks using $ID$ verification, where the attacker broadcasts fabricated cell ID and distributes fabricated public keys.  
The attack fails because the user equipment verifies the base station ID with the certificates of the registered base station database stored in a blockchain ledger. Our usage of permissioned blockchain prevents the attacker from distributing fabricated public keys because only the core network can write into the blockchain ledger.


\subsubsection{Against Base Station Spoofing} In the base station spoofing attack scenario, the malicious base station fails to prove the authenticity of its SIB1 messages. We test two cases for the malicious base station: (a) it takes a legitimate SIB1 message but includes a digital signature with its private key, and (b) it captures a legitimate digital signature but fabricates the SIB1 message. In both cases, SIB1 authentication fails because a valid digital signature can only be generated from a legitimate SIB1 message with a legitimate base station private key. The public key in the base station certificate is sufficient to thwart the base station spoofing attack. However, we authenticate the base station using only the SIB1 message, if the attacker exploits vulnerabilities of other RRC messages, then our scheme cannot provide complete defense against it, as we mentioned in our comprehensive threat analyses (section~\ref{subsec:threat_analyses}).

\begin{figure}
\centering

\begin{subfigure}{0.459\columnwidth} 
    \centering
    \includegraphics[width=1\textwidth]{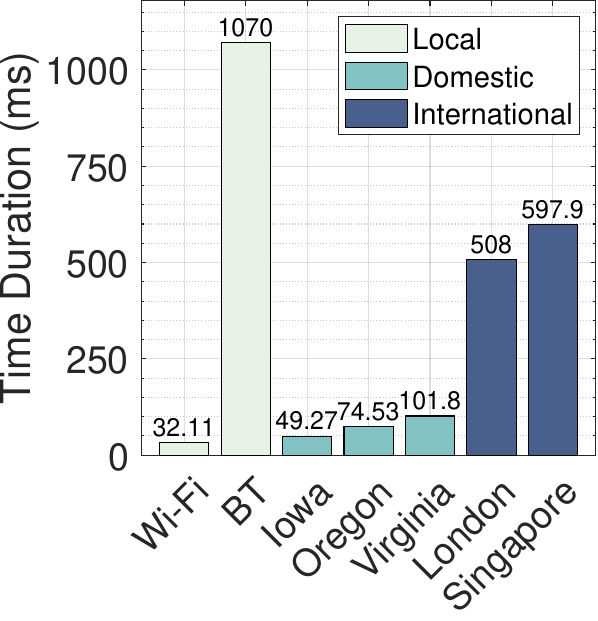}
    \caption{Certificate delivery time.}
    \label{fig:certDeliveryTime}
\end{subfigure}
\quad
\begin{subfigure}{0.459\columnwidth} 
    \centering
    \includegraphics[width=0.96\textwidth]{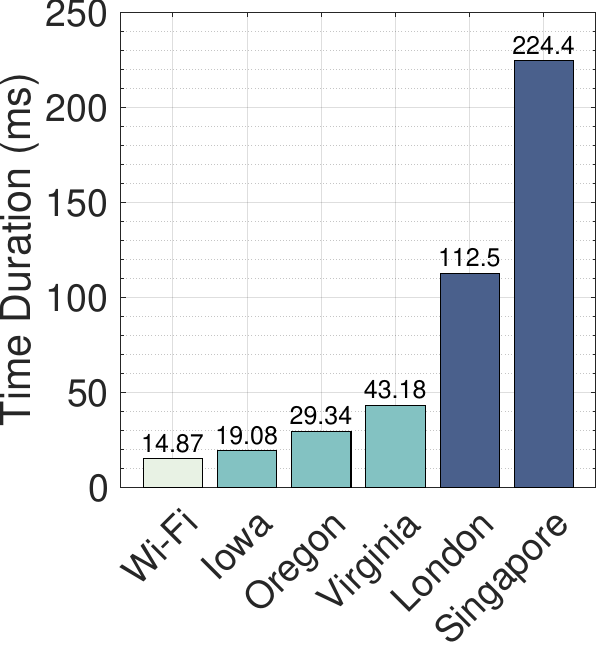}
    \vspace{0.7px}
    \caption{Transmission delay in RTT.}
    \label{fig:pingRoundTripTime}
\end{subfigure}
\caption{Offline blockchain-based certificate delivery performances when varying the certificate database source locations (Local vs. Cloud) and the communication protocols (Wi-Fi vs. Bluetooth). 
\vspace{-8px}}
\label{fig:sizes}
\end{figure}


\subsubsection{Against Communication Injection, Including Wormhole}
As illustrated in Figure~\ref{fig:attacK_wormhole}, we implement the wormhole attack by capturing a legitimate SIB1 message with the corresponding digital signature at one location and broadcasting it in another location. Although the scheme successfully verifies the SIB1 message, the $L$ verification fails because the user equipment senses its location to validate the distance to the base station. 
In such a case, the measured distance is beyond the threshold selected by the user equipment. The location information in the base station certificates, and user equipment location provide spatial integrity to prevent such attacks. 



In our implementation, we also capture a legitimate SIB1 message with the corresponding digital signature and replay the message to the benign user equipment as illustrated in Figure~\ref{fig:attacK_replay}. In such a scenario, the user equipment in our scheme invalidates the SIB1 message using $t$ verification because the SIB1 message reception time is higher than the control threshold. The timestamp $t$ (described in Section~\ref{subsec:authentication_factors}) provides temporal integrity of the SIB1 message, thus preventing wormhole and man-in-the-middle attacks.  

\begin{figure}
\begin{subfigure}{0.46\columnwidth}
    \centering
    \includegraphics[width=1\textwidth]{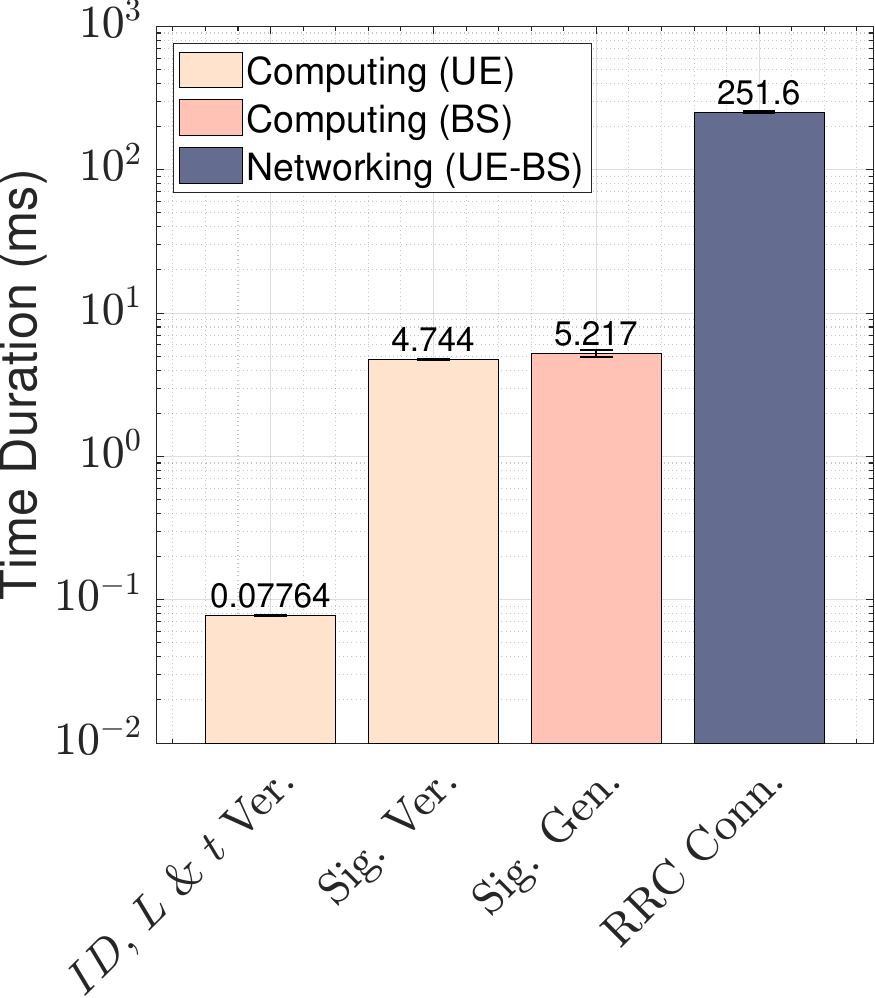}
    \caption{Time durations 
    on the user equipment and base station.}
    \label{fig:online_performance_computation}
\end{subfigure}
\quad
\begin{subfigure}{0.45\columnwidth}
    \centering
    \includegraphics[width=1\textwidth]{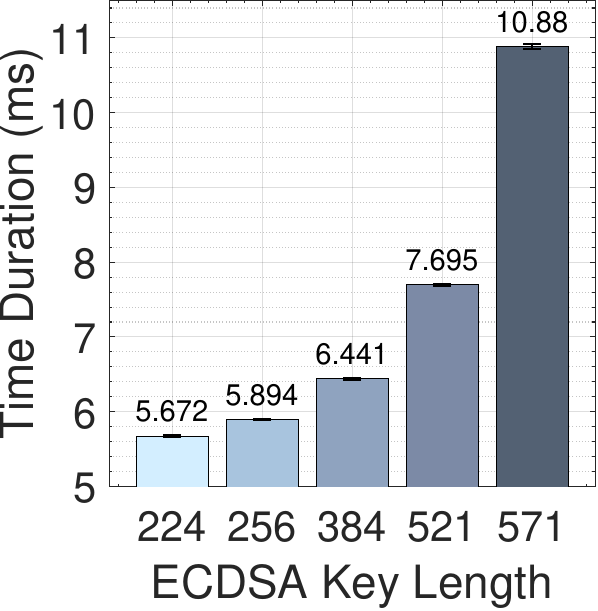}
    \vspace{1.2px}
    \caption{Signature verification times varying ECDSA.} 
    \label{fig:sign_verification_algorithm}
\end{subfigure}
\quad
\caption{Our scheme's online performance.}
\label{fig:sizes}
\end{figure}

\vspace{-1px}
\subsection{Offline Results for Blockchain Key and Certificate Delivery}
\label{subsec:offline_results}
We implement the offline blockchain for the base station certificate delivery and run experiments to measure the blockchain size, blockchain scalability requirement (the number of certificates per second required to support the 5G applications), and the certificate delivery time. As described in Section~\ref{subsec:our_scheme_blockchain}, the core network generates the certificates (its write control is implemented in the smart contract), while the blockchain can be read and stored by other nodes including the user equipment. As discussed in Section~\ref{sec:experimental_setup}, we experiment in both local ad hoc and remote Internet-based environments and vary the communication protocols to highlight the blockchain's strength in automatic synchronization independent of the communication protocols (as discussed in Section~\ref{subsec:our_scheme_blockchain}). We also analyze the scalability requirement (which is a critical challenge identified in the general blockchain and cryptocurrency research) and compare it with other blockchain applications to establish the blockchain's appropriateness for our application in this section.

\subsubsection{Blockchain Scalability}
In the blockchain use for base station certificate delivery, each new transaction or record gets updated when there is a change in the base station registration, including a new base station construction or a public-key revocation, which is far less frequent than the scalability requirements of other current blockchain implementations and practices such as those for financial transactions. 
Our scheme requires only 0.0266 certificate/transactions per second, which is three orders of magnitude smaller than the other blockchain applications for cryptocurrency such as Ethereum and Bitcoin. 

\subsubsection{Offline Time Overhead Performance}
We use diverse communication channels (Wi-Fi/BT) and sources (local/cloud) to demonstrate the robustness of blockchain-based certificate delivery to the user equipment. Figure~\ref{fig:certDeliveryTime} shows our experimental results for certificate delivery (34680 Bytes) from the 3rd node and the cloud server using ad hoc and infrastructure networks. The ad hoc network connectivity using Wi-Fi provides the fastest certificate delivery (32.22 ms) from the 3rd party node as compared to other locations and connectivity. The certificate transfer using ad hoc Bluetooth takes the longest time (1070.2 ms), which is natural due to the physical-layer design of Bluetooth providing low data rate/bandwidth. We also measure the certificate delivery from the remote cloud servers, connected to the user via the Internet (TCP/IP) and through the 3rd Party Node as the last networking hop via Wi-Fi. As shown in Figure~\ref{fig:pingRoundTripTime}, the RTT time duration increases as the remote node to initiate the blockchain transmission is located geographically farther. 
The cloud server in Iowa provides the lowest certificate delivery time (52.23 ms) and the cloud server in Singapore takes the highest time (1060.05 ms or 1.06 seconds) to deliver the certificate. 
While these measurements can be high and can reach up to about 1 second, the offline cost in the time overhead is significantly cheaper than the real-time online cost in Section~\ref{subsec:results_online}.

\begin{figure}[t!]
\begin{subfigure}{0.421\columnwidth}
    \centering
    \includegraphics[width=1\textwidth]{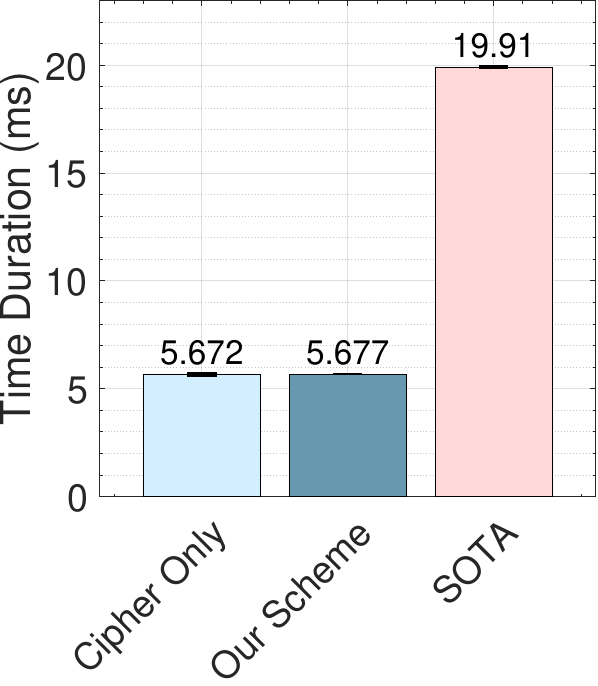}
    \caption{Signature verification time durations at user equipment.} 
    \label{fig:sign_verification_comparison}
\end{subfigure}
\quad
\begin{subfigure}{0.525\columnwidth}
    \centering
    \includegraphics[width=1\textwidth]{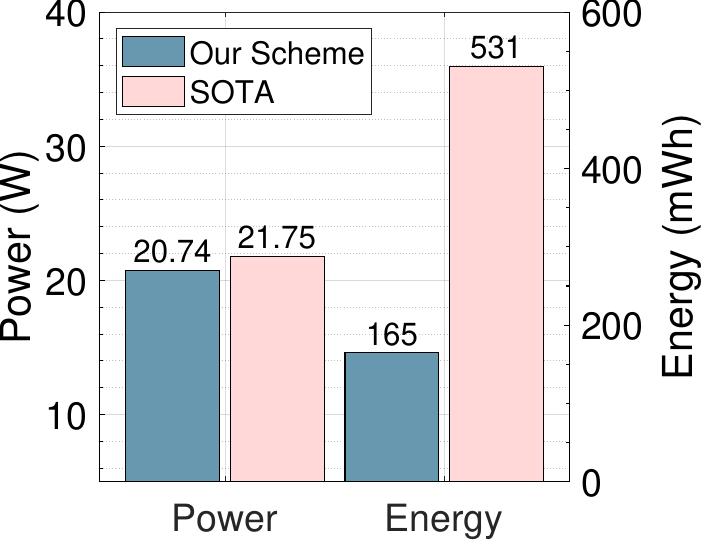}
    \vspace{6px}
    \caption{Power (measured) and energy (calculated) consumption at the user equipment.}
    \label{fig:energy}
\end{subfigure}

\caption{Our scheme vs. SOTA comparison.\vspace{-5px}}
\label{fig:sizes}
\end{figure}
\vspace{-0px}
\subsection{Online Results for multi-factor Authentication} 
\label{subsec:results_online}

We compare our scheme against the SOTA described in Section~\ref{subsec:sta_r} in the computational time, energy consumption, and the RRC connection setup overheads. We also vary the user equipment locations for radio experiments and measure the signal-to-noise ratio (SNR) and its impact on the RRC connection overheads.

\subsubsection{Our Scheme's Online Performances}
We take a systems approach to measure the performance across the online operations of our scheme. 
In Figure~\ref{fig:online_performance_computation}, we show that $ID, L, t$ verification time (0.0776) and signature verification time (4.744 ms) at the user equipment are much lower than the RRC connection establishment time duration between the user equipment and the base station (251.6 ms). The user equipment computing is only 1.89\% compared to the entire RRC networking between the user equipment and base station. Also, we add only a total of $72$ bytes of overhead in the SIB1 message using the ECDSA-224 algorithm (signature - 64B, nonce - 4B, and timestamp - 4B).

Our scheme enables the use of the ECDSA ciphers with greater keys and security strengths, while the use of the ECDSA with greater keys is prohibited in SOTA as described in Section~\ref{subsubsec:greater_security}. We experimentally verify this and measure the time overheads when varying the ECDSA key lengths in Figure~\ref{fig:sign_verification_algorithm} to show that the time overheads increase as the key lengths increase. In terms of SIB1 byte length overhead with greater ECDSA key length, only the signature size contributes to SIB1, which is below the maximum SIB1 size (376 bytes).

\subsubsection{Our Scheme vs. SOTA}
\label{subsubsec:os_vs_sota}
We focus on our scheme implementation using ECDSA-224 when comparing our scheme vs. SOTA because SOTA does not support the other ECDSA digital signature algorithms as discussed in Section~\ref{subsubsec:greater_security}. 
Figure~\ref{fig:sign_verification_comparison} compares the computing overhead of the signature verification for our scheme vs. the SOTA. 
Because the SOTA requires three signature verifications on the user equipment due to certificate chaining, while our scheme only requires one, the signature verification overhead of the SOTA (19.91 ms) is 3.50 times greater than our scheme (5.68 ms). Our scheme's signature verification after incorporating it within srsRAN (5.68ms) is close to the digital signature verification without the srsRAN incorporation (5.67ms). 

The signature verification efficiency of our scheme compared to the SOTA results in the energy efficiency of our scheme, which is important for battery-equipped and energy-constrained user equipment. We take the power measurements in our experiment and then calculate the energy consumption by multiplying the average power consumption and the average time duration for the signature verification. 
As seen in Figure~\ref{fig:energy}, the SOTA takes 3.22 times more energy (531.38 milliwatt-hour versus 165.21 milliwatt-hour) than our scheme to verify the signatures, because our scheme verifies the digital signature once while the SOTA verifies three digital signatures to complete the authentication. 


\subsubsection{Wireless experimentation}
\label{apndx:radio_experiments}

\begin{figure}[t!]
\centering
\includegraphics[width=0.76\columnwidth]{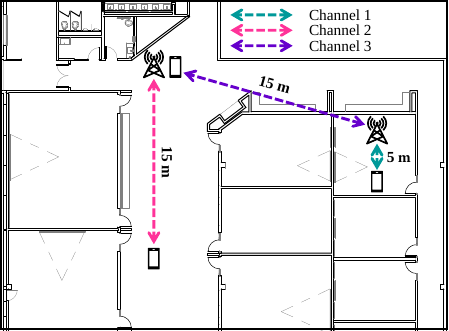}
\caption{
The base station and the mobile user locations for the online channel experiments.}
\label{fig:building_layout_image}
\end{figure}

\begin{figure}[t!]
    \begin{subfigure}{0.464\columnwidth}
        \centering
        \includegraphics[width=1\textwidth]{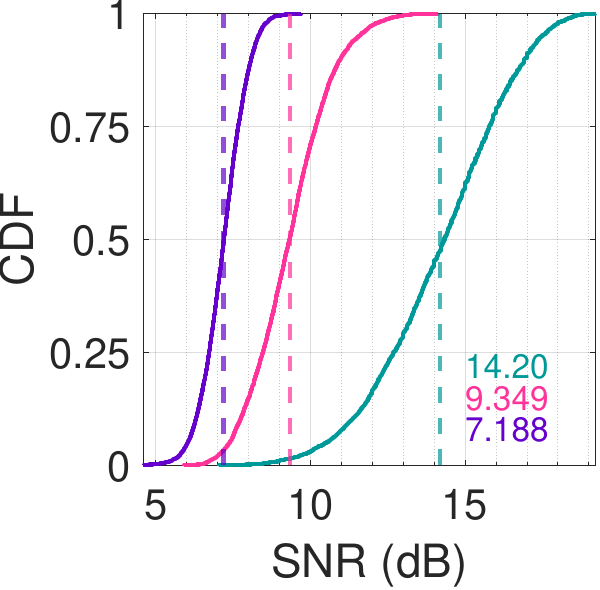}
        \caption{Signal-to-noise ratio (SNR) for each channel.}
        \label{fig:SNR_vary_location}
    \end{subfigure}
    \label{fig:rrc_snr}
    \quad
    \begin{subfigure}{0.48\columnwidth}
        \centering
        \includegraphics[width=1\textwidth]{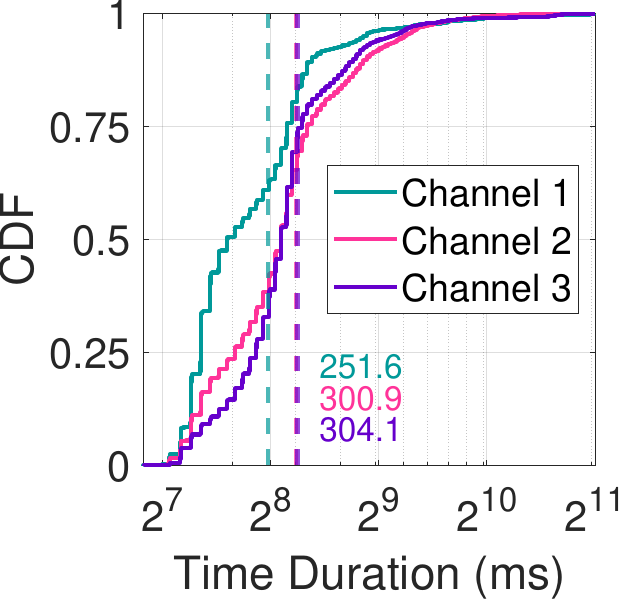}
        \caption{\vspace{-1px}
        RRC connection setup time duration for each channel.}
        \label{fig:RRC_time_vary_location}
    \end{subfigure}
    \caption{
    Cumulative distribution functions for the RRC connection time and radio channel condition, where the dashed line denotes the average of each channel.}
\end{figure}

We vary the wireless channel conditions between the mobile user and the base station, as shown in Fig.~\ref{fig:building_layout_image}. Channel 1 has the shortest distance of $d=5$ meters, while Channels 2 and 3 have the same distance of $d=15$ meters. Channel 2, however, has a line-of-sight (LOS) channel while Channel 3 has physical barriers in walls between the mobile user and the base station (non-line-of-sight or NLOS). 

Because shorter distance (smaller $d$) and LOS channel provide better wireless channels, Channel 1's signal-to-noise ratio (SNR) performance is better than Channel 2's, and Channel 2's is better than Channel 3's (14.2 dB vs. 9.35 dB vs. 7.19 dB) as shown in Fig.~\ref{fig:SNR_vary_location}. The channel conditions also affect the RRC connection setup time, and the better the channel quality, the lower the connection time duration. Channel 1 takes 251.58 ms for the RRC connection setup, while Channel 2 takes 300.90 ms and Channel 3 takes 304.11 ms, as shown in Fig.~\ref{fig:RRC_time_vary_location}.

\vspace{-1px}
\vspace{-4px}
\section{Conclusion}
\label{sec:conclusion}
To defend against fake or malicious base stations, we introduce a base station certificate that includes its ID, public key, and location information, then deliver the certificate from the network to the user equipment using blockchain to enable multi-factor base station authentication at the user equipment. 
We use offline communication for base station certificate construction and delivery while using online communication for multi-factor base station authentication which involves sequential verifications.   
We provide comprehensive threat and defense analyses in current practice, SOTA, and our scheme to motivate and inform the certificate construction and use the certificate fields for different factors in multi-factor authentication. 
Our work advances security and performance over the SOTA to enable multi-factor authentication using multiple orthogonal channels on the user equipment, to defend against wormhole using the location information, and to enable greater performance and security strengths. 
Our scheme involving a novel base station certificate and its blockchain-based delivery and use in 5G RRC enables stronger security in ECDSA cipher with greater key length. Our scheme also provides more than three times better performance advantages in computational time and energy consumption by verifying the digital signature only once, unlike SOTA. We experimentally validate the security and performance advantages of our scheme over SOTA by implementing it using open-source software (srsRAN, Ethereum, OpenSSL) and SDR. 
We expect our research to inform and guide to secure the design of the base station communication protocol in the future-generation 6G and beyond-5G. 
\vspace{-1px}

\section*{Acknowledgment}
This work was supported in part 
by National Science Foundation under Grant No. 1922410. 

\bibliographystyle{IEEEtran}
\bibliography{bibliography}



\end{document}